\documentclass[12pt]{article}
\usepackage{amsmath}
\usepackage{chicago}
\usepackage[left=3cm,right=3cm,top=3cm,bottom=3cm]{geometry}
\usepackage{mathpazo}
\usepackage{tabularx}
\usepackage{graphicx}
\usepackage{graphics,subfigure}
\usepackage{hyperref}
\usepackage{booktabs}
\usepackage{lscape}

\usepackage{graphicx,epstopdf} 
\newcommand{\figref}[1]{\hyperref[#1]{Figure~\ref*{#1}}}
\newcommand{\tabref}[1]{\hyperref[#1]{Table~\ref*{#1}}}

\begin{document}

\title{The welfare effects of unemployment insurance \\ in Argentina \\ \large{New estimates using changes in the schedule of transfers}}

\author{Martin Gonzalez-Rozada \\ {\small UTDT} \and Hern\'{a}n Ruffo\thanks{We would like to thank the ``Direcci\'{o}n General de Estudios Macroecon\'omicos y Estad\'{i}sticas
\mbox{Laborales''} from the ``Secretar\'{i}a de Gobierno de Trabajo y Empleo del Ministerio de Producci\'{o}n
y Trabajo'' for granting us access to the data and processing our codes, making this work possible. We are especially grateful to Bernardo D\'{i}az de Astarloa, Juan Zabala, and Mar\'{i}a Victoria Castillo-Videla. } \\
{\small UTDT}}

\date{February 2022}

\maketitle
\setcounter{page}{0} \thispagestyle{empty}

\vspace{-1.5cm}
\begin{abstract}
Unemployment insurance transfers should balance the provision of consumption to the unemployed with the disincentive effects on the search behavior. Developing countries face the additional challenge of informality. Workers can choose to hide their employment state and labor income in informal jobs, an additional form of moral hazard. To provide evidence about the effects of this policy in a country affected by informality we exploit kinks in the schedule of transfers in Argentina. Our results suggest that higher benefits induce moderate behavioral responses in job-finding rates and increase re-employment wages.
We use a sufficient statistics formula from a model with random wage offers and we calibrate it with our estimates. We show that welfare could rise substantially if benefits were increased in Argentina. Importantly, our conclusion is relevant for the median eligible worker that is strongly affected by informality.

\vspace{0.3cm}

 \vspace{0.3cm}
\end{abstract}

\clearpage

\section{Introduction\label{S:Intro}}

The existence of informality restricts the governments in developing countries when designing policies and transfers. This is particularly true for unemployment insurance (UI). 
Informal jobs are unobserved by the government; workers can collect UI and informal labor income at the same time. This unobservable job is a source of moral hazard that adds to the traditional one that arises due to unobserved search effort. The actual welfare effects of UI, then, can be affected by informality. The fact that many developing countries lack any kind of unemployment insurance suggests that informality could seriously affect the ability of the government to provide transfers.


The evaluation of the optimality of unemployment insurance (UI) has benefited greatly from the sufficient statistics approach.
This method uses theory to identify the welfare effects of a given policy, deriving a formula that summarizes marginal welfare gains and marginal social costs of a reform. These welfare effects are written as functions of a few elasticities, which are identified as the sufficient statistics that summarize the main welfare effects. Then, empirical methods from the program evaluation literature can be used to estimate these elasticities \cite{Chetty-Finkelstein,Chetty09SuffStats}.



Several papers have contributed in this tradition. Many of them concentrate on the level of UI \shortcite{Chetty08,CardChettyWeber,ShimerWerning07,LandaisMichaillatSaez18ap}; some refer to the optimal extension of UI \shortcite{Schmiederetal12,gerard-gonzaga,Nekoei-Weber};
and some evaluates both the level and potential duration \shortcite{Landais15,GRRuffo2016}.

This paper follows the sufficient statistics literature to provide an evaluation of the UI system in Argentina, a country affected by high informality. Following \citeN{ShimerWerning07}, we use a McCall search model of random wage offers to analyze the welfare effects of UI level. In this context, the worker has a reservation wage policy and accepts any job for which the wage  is higher than the reservation wage. Then, under some conditions, the reservation wage, net of taxes, is a sufficient statistic of the worker's welfare. Any increase in the net reservation wage would be evidence of a welfare improvement among the unemployed. While reservation wages are unobservable, the change in re-employment wages can be used to infer the effects on reservation wages. At the same time, the required change in taxes can be estimated through the effects of UI transfers on the government budget.

After providing an optimality formula we calibrate it using administrative data from Argentina from 2005 to 2009.
In this country, the UI transfer is determined as a replacement rate, but with a minimum and a maximum level. In particular, the maximum level is restrictive, affecting the median beneficiary. These restrictions generate kinks in the schedule of UI with respect to a reference wage. Exploiting these discontinuities in the determination of the benefit level, we identify the effects of UI on several outcomes implementing regression kink design (RKD).

We find that a marginal increase in benefits rise re-employment wages, suggesting that the effect on reservation wage is substantial. Additionally, we measure the effect of this marginal rise on the government budget, including both the mechanical effect, ie. higher transfers to the same spells, and the behavioral responses, ie. longer unemployment durations. We find that the effects on the government budget are small, evidencing a very low behavioral response to UI level.

When we take the point estimates to the optimality formula we conclude that welfare would increase with a more generous UI transfer. The welfare gains are substantial and measured with relatively precise estimates. We find that welfare gains are between 7 and 21 cents per each monetary unit of UI increment. Considering that our implementation underestimates the net welfare gains of UI, this positive effects are remarkable.

We then explore whether our results are robust to a number of extensions. We first consider subsamples of workers, dividing them by age and pre-unemployment characteristics, making each sample more homogeneous. Then we use different RKD methods, including sharp and fuzzy RKD, and different bandwidths, including MSERD-optimal, FG, and other bandwidths. Furthermore, we provide additional estimates identifying the effect with data before and after a change in the schedule, that occurred in 2006. In that year, the maximum and minimum levels of UI were increased. We show that our conclusion, that the level of UI should be increased in Argentina, is robust to variety of samples and methods.


This paper relates to the long literature that estimates the effects of unemployment insurance on finding rate of the unemployed. Examples of these papers are \citeN{meyer90}, \citeN{Katzmeyer90}, \citeN{Machin99}, \citeN{ArellanoBoverBentolila} among others.
Usually, this literature have found significant and negative effect of UI on finding rates of the unemployed, increasing the duration in unemployment state. The results on the quality of re-employment jobs and on re-employment wages are far less frequent, and usually the literature have not found any positive effects of UI \cite{CardChettyWeber,Lalive07AER,Schmieder-vonWachter-Bender16AER}. A few papers have found positive effects of UI level or extension on wages \cite{CentenoNovo09,Nekoei-Weber}. Our paper contributes to this particular literature by showing positive and significant results of the effects of UI level on wages.

Our work connects to a particular branch of literature that have focused on identifying the effects of UI using the kinks in the schedule of UI transfers. Examples of this approach include \shortciteN{Cardetal2015}, \citeN{Landais15}, \shortciteN{camposetal}, \shortciteN{Leeetal2021}.

The remainder of the paper is as follows. In the next section we briefly describe an optimality formula that arises from a search model with random wage offers. In section \ref{S:System}, we present the characteristics of the unemployment insurance system in Argentina and we characterize the determination of the transfer level. In section \ref{S:Empirical} we present our empirical strategy and we describe our data. In section \ref{S:Results} we present our estimations of the effects of benefits on duration of spells and we evaluate the optimality formula.
Section \ref{S:Robustness} explores how the main conclusions change when we implement different empirical methods.
In section \ref{S:Conclusions} we conclude by discussing the interpretation of our results.

\section{The welfare effects of a higher UI transfer}\label{S:TModel}

In this section we use a search model to evaluate the welfare gains and costs of an increase in UI\ transfers. We base on a stationary random wage offer model \cite{Mccall70} and follow \citeN{ShimerWerning07} to derive a welfare formula that can be easily measured though the effects of UI on wages and on the costs of the policy.

The model describes the behavior of a risk averse worker
who is unemployed in period $t=0$. The worker consumes the period income and discounts the future at a rate $r$. Each period receives a wage offer at a rate $\lambda$. The wage offer is distributed according to the cumulative distribution function $F$. The worker continues in unemployment if the offer is rejected. If the worker accepts the offer, the worker becomes employed with labor income determined by a wage $w$ and a tax $\tau$. We consider the case of a constant separation rate, $\delta$, from employment to covered unemployment.

At the beginning of the unemployed spell the worker receives $b$ as a transfer from the UI system. The worker can loose the eligibility at a rate $\gamma$, in which case the worker receives a low unemployment assistance transfer, $b_a$, with unlimited duration. The worker also gets an informal (unobserved by the government) labor income $y$ if not formally employed.

The problem of the eligible worker is stationary and its solution is a reservation wage, $w^{\ast}$. This threshold determines the wage above which any wage offer is accepted. The analogous problem of the worker after UI exhaustion is also stationary and the corresponding reservation wage is lower than the one for the eligible.

In this setup, the eligible reservation wage, net of taxes, would be the monetary counterpart of the utility while eligible. Intuitively, the net reservation wage is the amount required by a worker to stop searching for a job and begin working. It is, thus, a measure of the welfare at the unemployment spell.

Taking this into account, a government aiming at maximizing the welfare of workers can use as objective function the after-tax reservation wage, $w^{\ast}-\tau$. Notice that both variables are functions of UI and that the reservation  wage is also a function of taxes.

Given this objective function, the partial derivative on the reservation wage with respect to UI level would represent the welfare gain of the government transfers. The change of re-employment wages would, thus, provide evidence of this effect. At the same time, any increase in UI would require an increase in taxes, both because of the higher expenditures in UI transfers and the increase in the time spent in unemployment due to the changes in the incentives to search. The rise in taxes implies a welfare loss that could be measured empirically through an estimate of the additional expenditure.

The UI would be in its optimal level whenever these marginal welfare gains and marginal welfare losses exactly balance.
This condition can be approximated by
\begin{equation}\label{eq:FOC1}
\frac{\partial w^{\ast}}{\partial b} = \frac{\tau^{\prime}}{1+\tau^{\prime}}
\end{equation}
where this formula is based on the assumption that $\frac{\partial w^{\ast}}{\partial b}=\frac{\partial w^{\ast}}{\partial \tau}$, as in the CARA case, and where $\tau^{\prime}$ represent the total change in taxes when benefits are increased. While in this presentation we impose particular assumptions to get the formula in (\ref{eq:FOC1}), in Appendix \ref{S:AppModel} we argue that this condition will still be valid in a more general context and that can be considered as an approximation in a wide range of cases. We also describe there that the equation can be written as

\begin{equation}\label{eq:FOCf}
\frac{\partial {\overline{\log w}} }{\partial b} w^{\ast}\times \Lambda= \frac{\frac{\partial R}{\partial b}} {\frac{1}{\delta}+R/b}
\end{equation}
where $\overline{\log w}$ is the mean of log accepted wages, $R \equiv \sum_{d} S_d b_{d}$ is the expected amount of benefits paid to the eligible worker at the beginning of UI, where $S_d$ is the expected survival rate up to duration $d$ and where $b_d$ is the benefit paid in duration $d$. In the above equation,
 $\Lambda \equiv \frac{\sigma^2}{\sigma^2-var}$. For this approximation the wage offer  is assumed to distribute log-normal with parameters $\mu$ and $\sigma$, and $var$ is the variance of the log accepted wages.

Equation (\ref{eq:FOCf}) shows that to characterize the optimality of the policy two effects of UI level should be measured: the effect on the log of re-employment wages and the effect over the benefits paid, that is, the derivatives in the equation.
In addition to these partial effects of benefits, several parameters should be calculated, including the separation rate, $\delta$, the expected amount of benefits paid to eligible workers, $R$, the reservation wage level, $w^{\ast}$, and the relevant parameters from the wage offer distribution to compute $\Lambda$. In Section \ref{S:Results} we show how we choose values for these parameters.

\section{Unemployment Insurance in Argentina}\label{S:System}

In this section we briefly describe the UI system in Argentina.\footnote{A more extensive description can be found in \citeN{GRRuffo2016} and in \citeN{GRRR}.}
UI was introduced in 1991 for private employees and was extended to construction workers in 2001. The program is financed by a 1.5 percent payroll tax on employers and is managed by both the Ministry of Labor and the National Social Security Administration (ANSES).
Eligibility requires that workers are laid off from their jobs due
to no fault of their own. The program covers all private sector employees except rural, domestic, school teachers, and university professors.

\tabref{tbl:tab2} shows the elegibility to UI
and its potential duration according to age group and the number of months of contributions to the
system before dismissal.
The rules apply to ``permanent workers''.\footnote{In this paper, we do not analyze seasonal contracts, which have a particular determination of
benefits because these UI beneficiaries represent less than 1\% of the total. We also abstract from the construction sector which has its own particularities in UI system.}

In Argentina, the UI benefit is paid at monthly basis. The initial UI transfer, $b_1$, is a fixed fraction, $\beta,$ of the pre-unemployment reference wage, $W_R$, which is computed as the highest pre-unemployment wage in the last six months. According to the rule, this fraction is set to 50\%.
There are, additionally, a minimum and a maximum benefit amount, $b_{L}$ and $b_{H}$ respectively. These levels were set to 150 and 300 ARS since the
1991, when the UI system was introduced. In March 2006 they increased by 100 ARS to 250 and 400 ARS, respectively. Thus, initial benefits are
\begin{equation}\label{eq:schedule}
b_{1}^{t} = \min \{ b^{t}_{H} , \max \{ b^{t}_{L} , \beta W_{R} \} \}
\end{equation}
where $t$ is the period, before or after March 2006.

Figure \ref{f:schedule} shows this schedule for the time span covered by our data. The grey line shows the function of $b_1^0$ on $W_R$ before the 2006 change. In that case, the minimum and maximum thresholds for the benefits determine two kinks in the relationship between the initial benefit and the reference wage. Let $k_H^0=b_H^0/\beta$ and $k_L^0=b_L^0/\beta$ identify the value of the reference wage at the high and low kinks, respectively. For the range $ k_L^0\leq W_R \leq  k_H^0$ this relationship is linear, but for $W_R\leq k_L^0$ and for $W_R\geq k_H^0$ the relationship is flat. The figure also plots the same relationship of $b_1^1$ and $W_R$, that is, for the period after the change in benefits.

Benefits also decrease over duration. Beneficiaries receive the full amount of UI during the first 4 months, 85\% of the full amount during months 5 to 8 and 70\% of the full amount from
9 to 18 months. Even under these reductions, the level of the benefit should satisfy to be higher than $b_L$.\footnote{The formula becomes in this case, $b_d^t= \max\{b_L^t, \rho_d \min\{ b_H^t, \beta W_R\}\}$, where  subscript $d$ stands for eligibility duration, and where  $\rho_d = 1$ if $d\leq4$, $\rho_d = 0.85$ if  $5 \leq d \leq 8$, $\rho_d = 0.70$ if  $d \geq 9$.}

\section{Empirical strategy}\label{S:Empirical}

\subsection{Methods}\label{S:Methods}

Consider the solid black line in \figref{f:schedule}. After the reform, an unemployed worker initially receives monthly benefit of $b^{1}_L$ if her reference wage, $W_R$, is such that half this wage is less than $b^{1}_L$. However, if half her reference wage is greater than $b^{1}_L$, then her benefit is a linear function of pre-unemployment wage until a maximum threshold of $b^{1}_H$. When half her reference wage is greater than $b^{1}_H$, the benefit received is remains constant at $b^{1}_H$. Therefore, the UI formula after the reform as a function of previous earnings has two kink points, at $k^{1}_L$ and $k^{1}_H$.

We use these kink points in the schedule of UI transfers to identify the causal effects of the level of benefits on outcome variables. We concentrate on one kink point, $k^{1}_H$, the high kink after the reform. There are several reasons for this. First, the high kink is closer to the median of the reference wage distribution, implying that it affects many workers and that there are more observations for any local estimation. Second, the high kinks affect the benefits paid during all the UI spell. On the contrary, the low kinks $k^{t}_L$ would affect only the first four months of benefits; after that duration, the fall in benefits would become binding and workers at both sides of the kink would have the same benefit level. Third, the sample associated to $k^1_H$ is larger in our data. Finally, those that begin their unemployment spell after April 2006 do not suffer any change in the UI system throughout their covered unemployment spell, what makes the exercise cleaner.
For these reasons we focus our estimations in the high kink $k^1_H$ and we continue the description of our method concentrating on this case.

The basic idea for identification is to look for an induced kink in the mapping between the assignment variable and the outcome variable that  coincides with the kink in the policy rule, and compare the relative magnitude of the two kinks.
This approach is called in the literature a regression kink design (RKD).

Consider the following constant-effect additive model,
\begin{equation}
Y = \alpha b + g(W_R) + \epsilon \label{eqoutput}
\end{equation}
In our setting, $Y$ is the outcome of interest (total benefits paid, re-employment wages), $b$ is the initial UI transfer and $W_R$ the running variable (pre-unemployment wage), and $\alpha$ is the parameter of interest, showing the effect of the benefit transfer on the outcome variable. Since UI benefit is a piecewise linear function of pre-unemployment wage, as in  equation (\ref{eq:schedule}), conditional on $W$ there is no variation in $b$ and the model is not identified in (\ref{eqoutput}). Nevertheless it may be possible to exploit the kink in the benefit rule to identify the causal effect of $b$ on $Y$. The idea is that if $b$ exerts a causal effect on $Y$, and there is a kink in the deterministic relation between $b$ and  $W_R$ at the kink point, $k$, then we should expect to see an induced kink in the relationship between $Y$ and $W_R$ at $k$.

\citeN{Nielsenetal} made precise the assumptions needed to identify the causal effects in (\ref{eqoutput}). They showed that if $g(W_R)$ and $E[\epsilon | W_R = k]$ have derivatives that are continuous in $W_R$ at the kink point, then
\begin{equation}
\alpha = \frac{\lim_{v_0 \rightarrow 0^+} \frac{d E(Y|V=v)}{d v}|_{v=v_0} - \lim_{v_0 \rightarrow 0^-} \frac{d E(Y|V=v)}{d v}|_{v=v_0}}{\lim_{v_0 \rightarrow 0^+} b(v_0) - \lim_{v_0 \rightarrow 0^-} b(v_0)} \label{eqsrkd}
\end{equation}
where $V= w - k$ is to normalize the kink point around zero. The expression on the right-hand side of equation (\ref{eqsrkd}) is simply the change in slope of the conditional expectation function $E[Y |V = v]$ at the kink point ($v = 0$), divided by the change in the slope of the deterministic assignment function $b(\cdot)$ at zero. Equation (\ref{eqsrkd}) is referred in the literature as the Sharp RKD estimand.

However, when the policy rule of interest depends on unobserved individual characteristics or is implemented with error then the error $\epsilon$ can be correlated with $W_R$ and equation (\ref{eqsrkd}) can give a non consistent estimation. \citeN{cardetal2012} show that to get a consistent estimation of the treatment effect on the outcome variable one replaces the known change in slope of the assignment rule at the kink with an estimate based on the observed data,
\begin{equation}
\alpha= \frac{\lim_{v_0 \rightarrow 0^+} \frac{d E(Y|V=v)}{d v}|_{v=v_0} - \lim_{v_0 \rightarrow 0^-} \frac{d E(Y|V=v)}{d v}|_{v=v_0}}{\lim_{v_0 \rightarrow 0^+} \frac{d E(b|V=v)}{d v}|_{v=v_0} - \lim_{v_0 \rightarrow 0^-} \frac{d E(b|V=v)}{d v}|_{v=v_0}}
\end{equation}
They called this a ``fuzzy regression kink design". \citeN{Dong2011} 
shows that the fuzzy RKD estimate can be recovered from an instrumental variable (IV) procedure.

\subsection{Empirical strategy}

We use this general equation that relates the output variable with the running variable,
\begin{equation}
E[Y|W_{R}=w]=\mu + \gamma_{1}\left( w-k\right) + \nu_{1}\left(w-k\right) D+X^{\prime}\beta
\end{equation}
for a bandwidth $|w-k|\leq h $ around the kink point, and where $D=1\left\{ w\geq k \right\}$ is an indicator variable for those observations above the kink point, $X$ is a vector of controls, $\mu, \gamma_{1}, \nu_{1}$ and $\beta$ are the parameters of the model. We are interested in the parameter $\nu_{1}$ which captures the change in the slope of the relationship between the outcome variable and the running variable at the kink point. Then, the sharp RKD estimated effect would be $\hat{\alpha}=\frac{\hat{\nu}_{1}}{\pi_{1}}$ where $\hat{\nu}_{1}$ is an OLS estimation of the previous equation, and where $\pi _{1}$ is the deterministic change in slope of $b$ at the kink. If necessary, an elasticity could be computed as
\begin{equation}\label{eq:elasticity}
\eta_{Y,b}=\hat{\alpha}\frac{b^1_{H}}{\bar{Y}}
\end{equation}
where $\bar{Y}$ is the mean of the outcome variable in the sample.

If the schedule for $b$ is not deterministic a fuzzy RKD has to be implemented. For this case we instrument $b$ with $(w-k) \times D$, and the IV estimate follows from the following reduced form and first
stage equations:\
\begin{eqnarray}
E[Y|W_{R}=w]=\mu + \gamma_{1}\left( w-k\right) + \nu_{1}\left(w-k\right) D+X^{\prime}\beta \label{eq:frkd}\\
E[b|W_{R}=w]=\lambda_{0} + \phi_{1}\left(w-k\right) +\pi_{1}\left(w-k\right) \times D+X^{\prime}\theta \nonumber
\end{eqnarray}
for a bandwidth $|w-k|\leq h$, and where the fuzzy RKD effect would be equivalent to $\hat{\alpha}^{RKD}=\frac{\hat{\nu}_{1}}{\hat{\pi}_{1}}$.

In both equations above it is assumed a linear effect of the running variable $W_R$. In section \ref{S:Robustness} we will also provide results using a quadratic function.

In our results, we prefer the \citeN{FG} (FG)  selection algorithm for the computation of the bandwidth. This method provides a wider bandwidth compared to alternatives considered and provides more robust and consistent results. Our choice is in line with the results in \citeN{cardetal2012}. They show that the FG bandwidth provides the lowest root mean square error in a Monte Carlo exercise. In any case, we also report the results for other bandwidths in Section \ref{S:Robustness}.

Several assumptions allow for the identification of the effect of $b$ by equation (\ref{eqsrkd}).
The first assumption states that the treatment variable, $b$, changes deterministically according to the rule stated above. This assumption can be tested by simple observation of the corresponding function in the data. It can also be lifted, as discussed above, by the fuzzy method, but in this case the relevance of the rule should be tested.
The second assumption relates to continuity. In particular, (i) the effect of the assignment variable on the outcome is smooth, (ii) any other determinant of the outcome change smoothly close to the kink, (iii) the marginal effect of the treatment variable on the outcome (if heterogeneous) is continuous close to the kink.
The third assumption is that there is no manipulation of the assignment variable or the treatment variable.

This last assumption seems credible in the context of our application. The rule is automatically implemented by the Social Security and monitored by the Ministry of Labor. It is not possible that workers could change their assignment variable, given that is declared by the employer. It is unlikely that an employer agrees to change the declared wage (for which should pay important contributions and severance pay) to surpass a given kink (for example). Furthermore, few people know the details of UI benefits when employed.
In Section \ref{S:Results} we provide important evidence about these three assumptions.

\subsection{Data}\label{S:Data}

Our data is based on the combination of several administrative databases.
These are (i) the monthly transfers to beneficiaries of the UI system (UB), (ii) the ``Sistema Integrado Previsional Argentino'' (SIPA); (iii) the monthly payments of independent workers to the Social Security system.
All of these data have identification numbers for workers, what allowed us to combine the databases and follow the same worker through the different situations: wage earner, self-employed and
UI beneficiary.
Our data, thus, follows each spell of covered unemployment and gathers information about the worker’s most recent job, pre-unemployment work history and re-employment job. We computed the duration of each spell as the difference in months between the period of the layoff and the period in which we first observe the worker as reemployed (either as a wage earner or self-employed).
Using all of the administrative sources, we construct variables of worker characteristics such as age, gender, number of children, etc.
We concentrate on beneficiaries of UI in 2005 through 2007. Several reforms were implemented during this period, whereas economic prospects and job creation were quite stable.

Column (1) of \tabref{tbl:characteristics} summarizes the main characteristics of the observations in the database restricting the sample to workers with reference wage between 75 and 4800 ARS.
The workers in our database are predominantly young males who earn relatively low wages in comparison average wages of formal workers.
About 50\% of the beneficiaries have a dependent spouse and the average number of dependent children is .77. In average, the workers contributed about 28 months of the last 36. Workers are eligible for approximately 9 months of UI. Many exhaust the funds they are entitled to, and on average eligible workers collect a total transfer of 2200 ARS in 7 months, with an average of 312 ARS per month.

In columns (2) to (4) the table shows UI\ beneficiaries in our pre-reform sample,  and divide them in three groups considering if the initial benefits are at the minimum, between the minimum and the maximum, and at the maximum level. The reference wages used to determine these groups are reported in the first line. Columns (5) to (7) report the analogous division for the post-reform sample.
The different groups defined by the level of reference wage show some differences in the predetermined variables, including age, tenure, contributions, proportion of males, among others. The average potential duration of UI also changes for these groups, increasing for those with higher reference wage. They also differ in outcomes: non-employment duration tend to decrease with the reference wage, average UI duration tend to increase (but less than UI eligibility). These changes in covariates do not invalidate our method, as long as they change smoothly around the kink, as the continuity assumption states.
Still, our results will include controls by age, sex, dependent spouse, dependent children, tenure in the last job, severance pay received,\footnote{Severance pay is not observed in the data. We use our computation of tenure, observed wages, and severance pay regulations to impute this variable for each observation.} the number of contributions and a set of fixed effects including eligibility for potential duration, as well as region, industry, year and month fixed effects. For wages, we add a set of dummies for non-employment duration (less than 5 months, 5 to 8, etc.). These intend to capture any difference in the level of re-employment wage due to different unemployment duration.

We study the effect of the level of benefits exploiting different kinks and defining for that purpose different samples. The sample for a top kink consists of workers with reference wage higher than the bottom kink plus a 5\%. To be clear, these thresholds define a sample and not the bandwidth, that is defined for each implementation of the RKD and computed through the FG method.  In all cases we restrict the analysis to workers displaced from indefinite contract with more than 12 contributions in the 36 months previous to unemployment.\footnote{We do this to avoid the possible change in eligibility criteria in 2006 (see Table \ref{tbl:tab2}).}

\section{Results}\label{S:Results}

\subsection{Graphical overview}\label{S:Graphs}

\begin{figure}[!h]
\caption{Schedule of initial UI transfer}\label{f:schedule}
  \vspace{-0.5cm}
\begin{center}
\includegraphics[trim = 90mm 100mm 150mm 40mm, width=4cm,height=7cm]{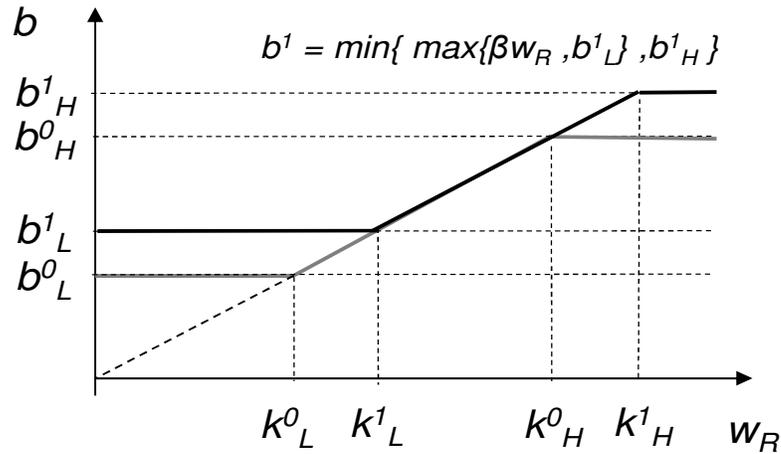}
\begin{minipage}{0.9\textwidth}\footnotesize
Note: The graph shows the schedule of the initial benefits, $b_{1}$, based on the pre-unemployment reference wage, $w_{R}$, up to a maximum and minimum level of transfer, $b_H$ and $b_L$.
The initial benefit schedule changed in March 2006, increasing the maximum and minimum levels in 100 ARS.
\end{minipage}
\end{center}
\end{figure}

\begin{figure}[!h]
\caption{Initial UI transfers}\label{f:schedule0}
\begin{center}

\begin{center}
\subfigure[2005 to March 2006]{\includegraphics*[trim = 30mm 155mm 51mm 26mm,clip,width=7.5cm,height=5cm]{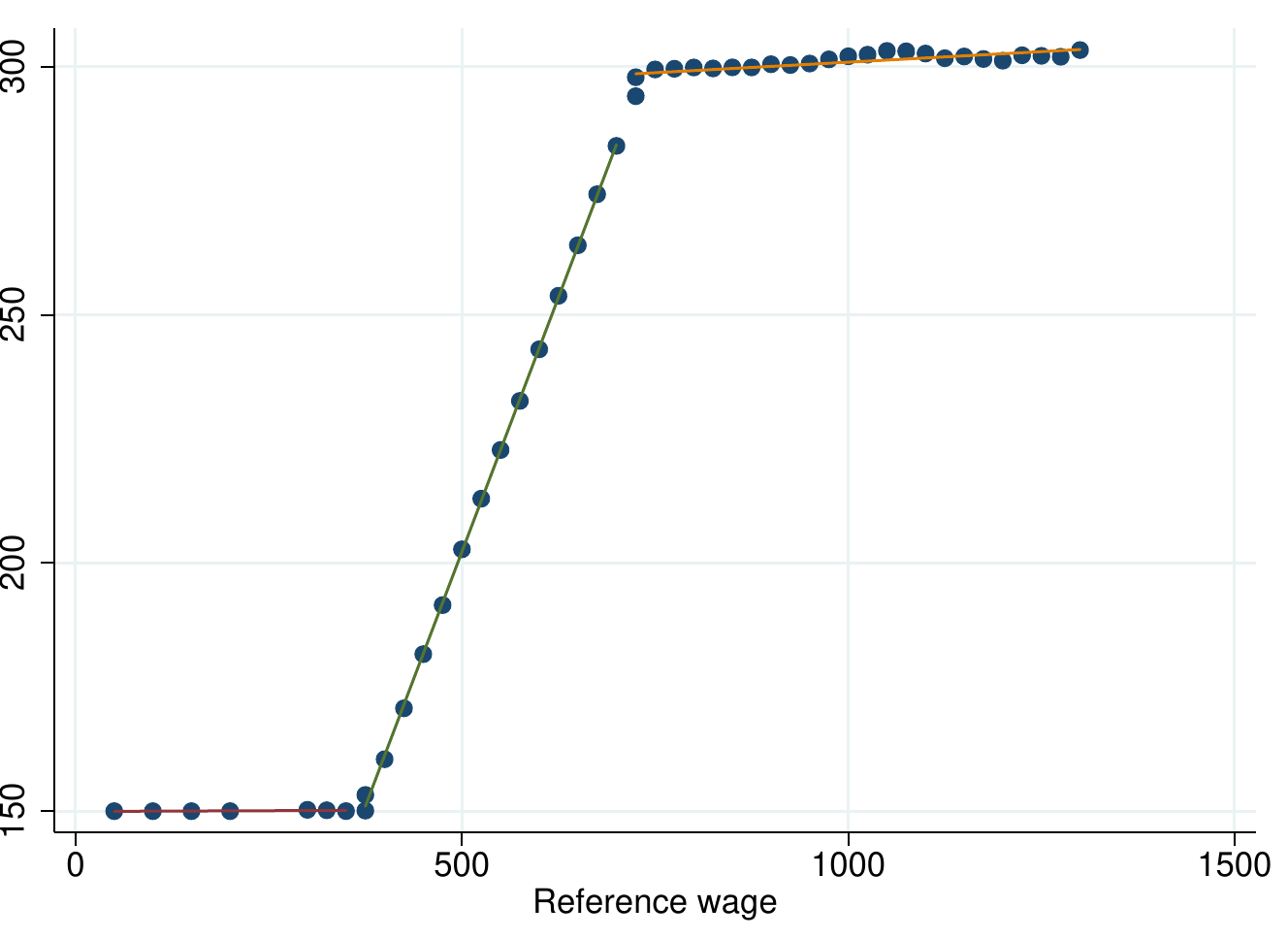}}
\subfigure[April 2006 to 2009]{\includegraphics*[trim = 30mm 155mm 51mm 26mm,clip,width=7.5cm,height=5cm]{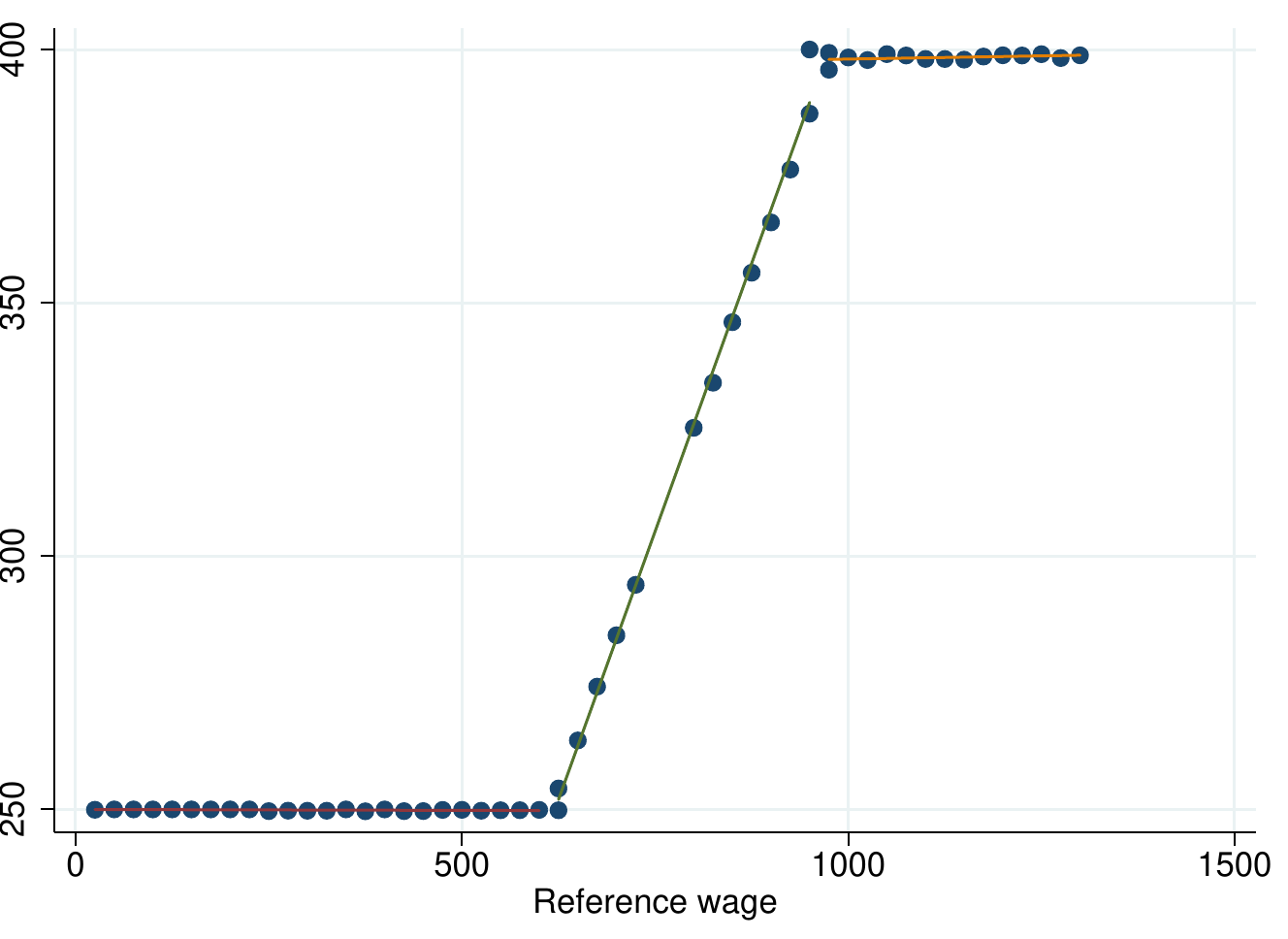}}
\end{center}

\begin{minipage}{0.9\textwidth}\footnotesize
Note: The graph shows the average benefit of the first month of UI within each bin of the assignment variable (gross pre-unemployment reference wage, in bins of 25 ARS). Lines are local linear regressions for three separate groups: those below the bottom kink, those above the top kink, and those in between kinks.
\end{minipage}
\end{center}
\end{figure}

\figref{f:schedule0} shows the first-stage relationship between reference wage and initial UI transfers. While some noise is present in the schedule of transfers, the overall shape is the same as the one presented in \figref{f:schedule}. This plot is constructed by dividing the assignment variable into bins and averaging out the initial UI benefit for each bin. Afterwards, regression lines are introduced for three samples: those below the bottom kink, those within both kinks and those observations above the top kink.

\begin{figure}[!h]
\caption{Total benefits paid in the UI spell}\label{f:Btot}
\begin{center}
\begin{center}
\subfigure[2005 to March 2006]{\includegraphics*[width=7.5cm,height=5cm]{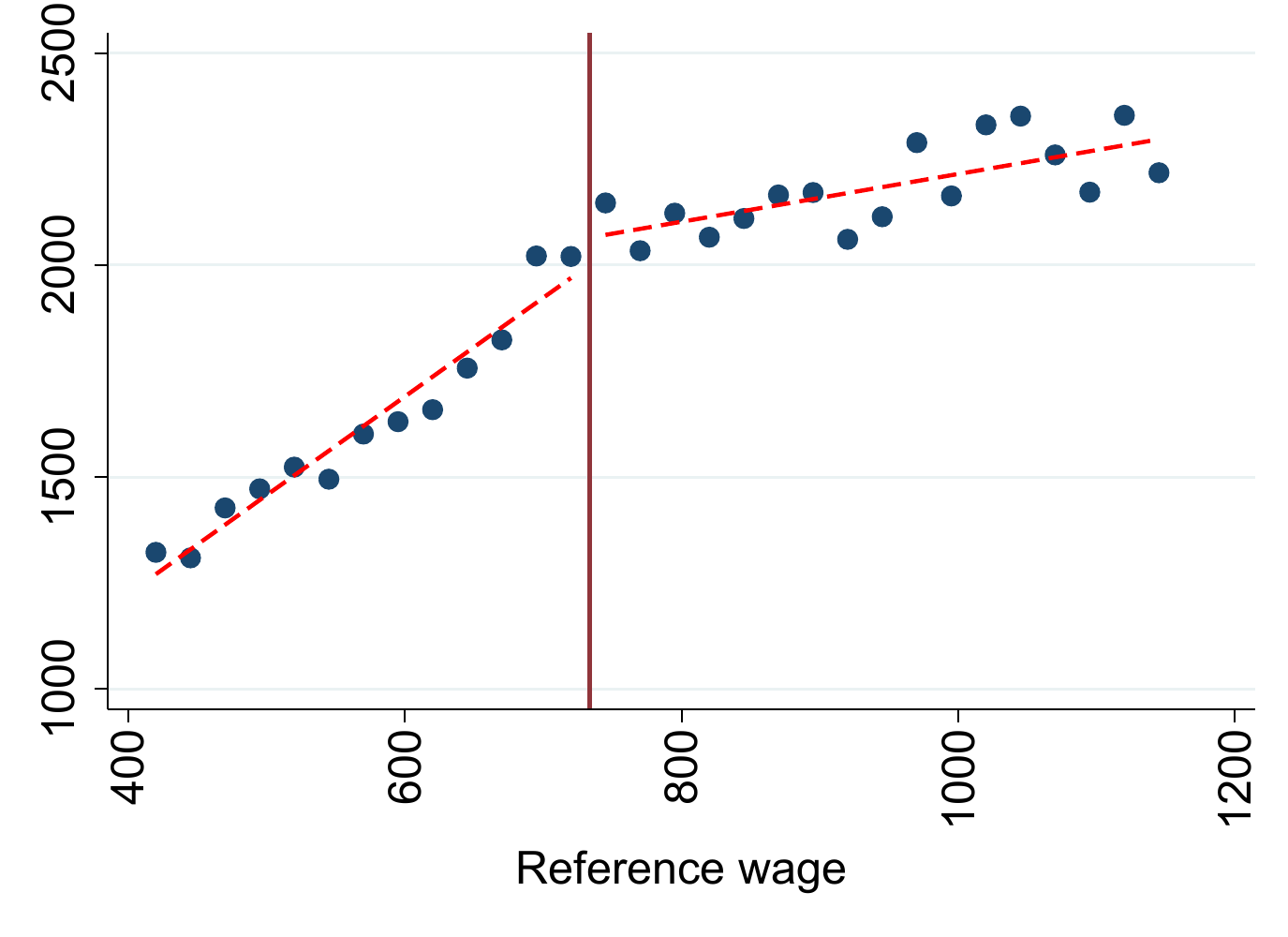}}
\subfigure[April 2006 to 2009]{\includegraphics*[width=7.5cm,height=5cm]{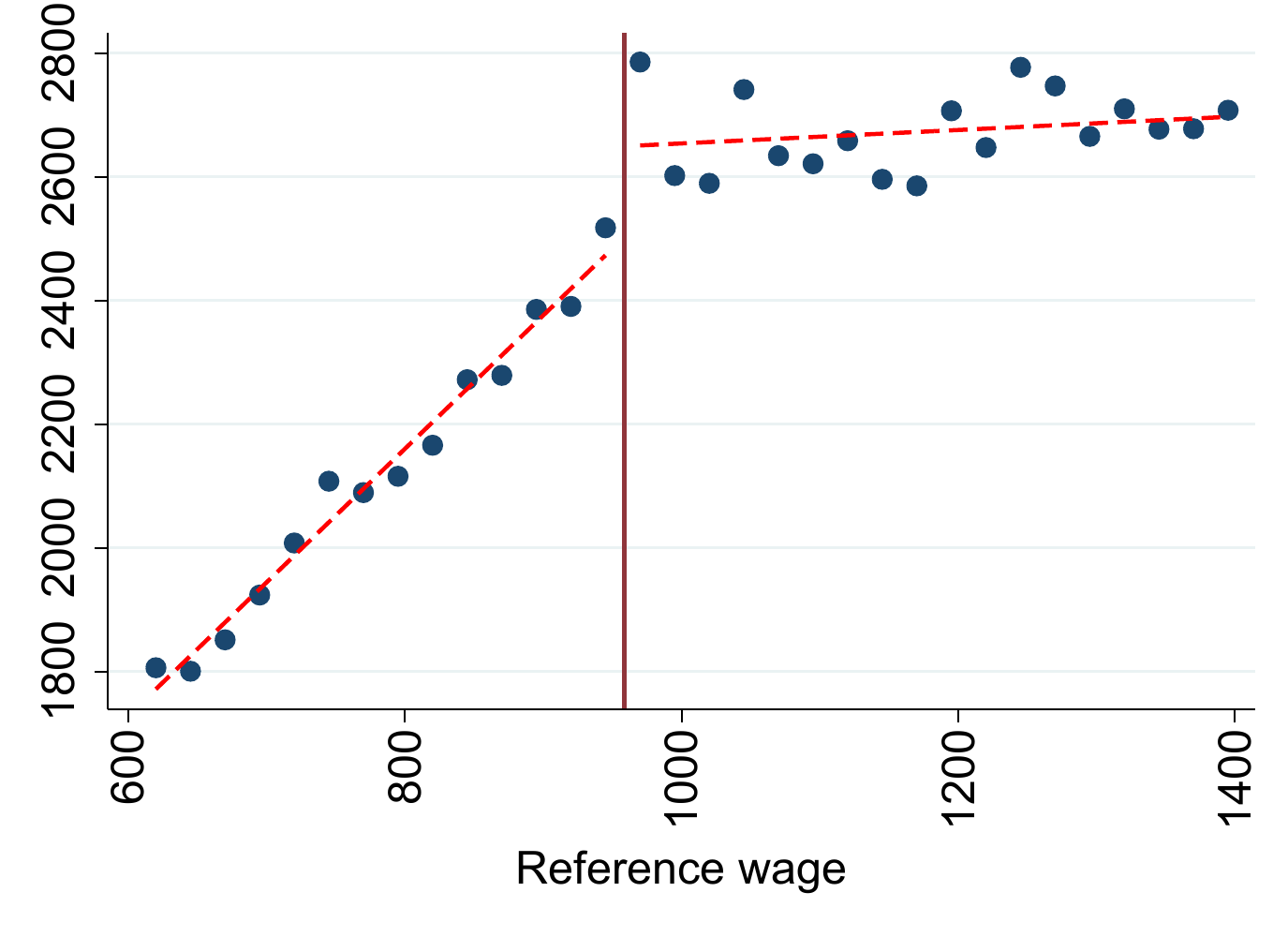}}
\end{center}

\begin{minipage}{0.9\textwidth}
\footnotesize
Note: The graph shows the average total UI transfer within the spell (in ARS) within each bin of the assignment variable (gross pre-unemployment reference wage). Lines show separate linear regressions for bins below and above the kink.

\end{minipage}
\end{center}
\end{figure}

\figref{f:Btot} shows analogous plots for the total amount of benefits paid, concentrating on the high kinks. This variable corresponds to $R$ in the model of Section \ref{S:TModel}. The kinks in these variables are clearly observable in both panels. Panel (b) is particularly clear.

There are two effects of the increase in benefits on the total amount paid that can be distinguished: a mechanical and a behavioral effect. The mechanical effect holds the number of collected benefits constant for all workers;\ any change in the level  of benefits would imply a proportional increase in the total UI\ paid.
The behavioral effect is the increase in the total UI paid due to changes in UI duration. This is the consequences of the lower job-finding rate, as a response to higher UI transfers.

\figref{f:Btot} includes these two effects, as well as changes in the composition in the characteristics of workers. If the mechanical effect was the only present, the change at the high kink would be similar to the initial schedule as in \figref{f:schedule}. The behavioral effect adds to this the change in the number of benefits collected. This effect  can be explained through our search model.  Consider the comparison of workers at both sides of the high kink. A lower effective replacement rate, at the right of the kink, could affect the job-finding probability for the worker. In such a case, the benefit duration could be affected by the kink in the benefits schedule. It is reasonable to expect a lower UI duration when replacement rate is reduced.
Besides these two effects, the total UI paid in the figure is also affected by the fact that potential duration changes (smoothly) with the reference wage.

\begin{figure}[!h]
\caption{Log of re-employment wages}\label{f:lnwage}
\begin{center}
\begin{center}
\subfigure[2005 to March 2006]{\includegraphics*[width=7.5cm,height=5cm]{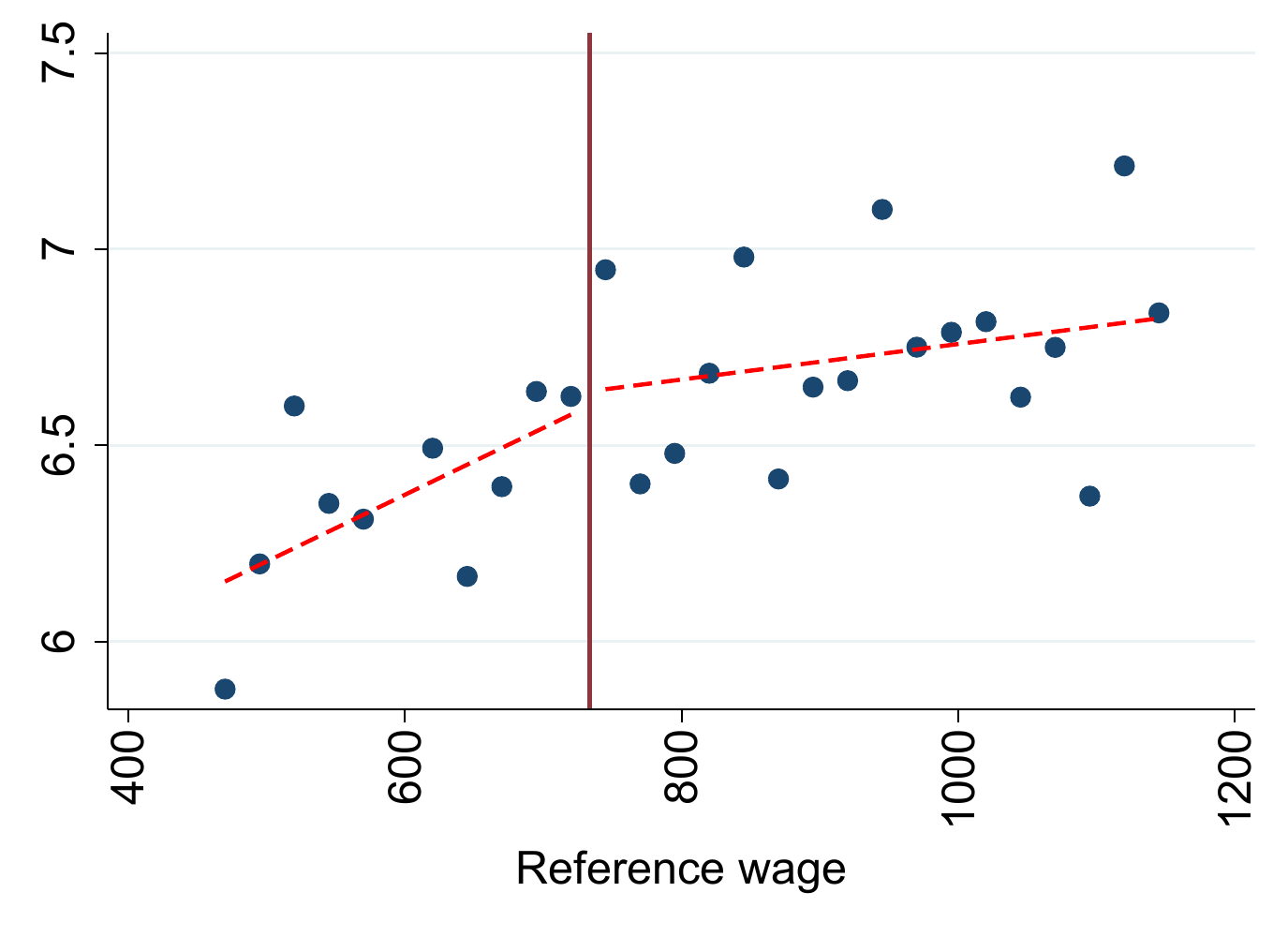}}
\subfigure[April 2006 to 2009]{\includegraphics*[width=7.5cm,height=5cm]{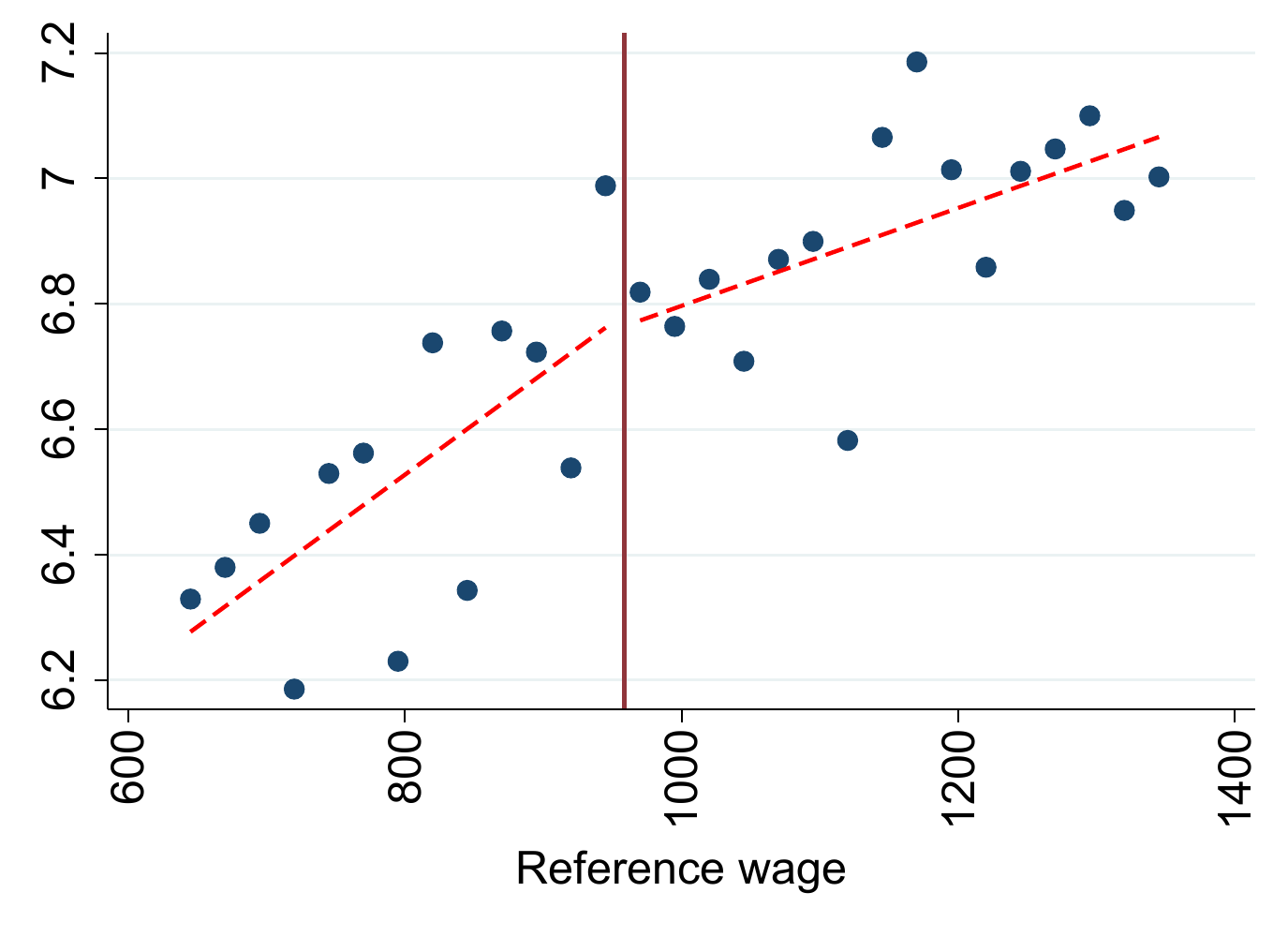}}
\end{center}

\begin{minipage}{0.9\textwidth}
\footnotesize
Note: The graph shows the log of re-employment wages within each bin of the assignment variable (gross pre-unemployment reference wage) trimming outliers. Lines show separate linear regressions for bins below and above the kink.
See notes to Figure \ref{f:Btot}.
\end{minipage}
\end{center}
\end{figure}

\figref{f:lnwage} presents the corresponding graph for the log of re-employment for the high kink. A mild change in slope is clear in both panels. Again, the fact that the slope is lower at the right of the kink is suggestive that the lower replacement rate affects negatively the re-employment wages.

These plots, then, suggest that UI seems to have effects on these outcome variables and that these effects can be identified through the RKD methods. Before proceeding to the estimations, we briefly revise the validity of using them.

A main assumption in RKD refers to the continuity of other determinants of the outcome variable around the kink and on the absence of manipulation of the assignment variable. We report evidence on these issues in what follows.

To account for the continuity assumption, we follow \shortciteN{cardetal2012} in constructing a predicted value of the outcome variable according to the relevant covariates. In particular, the predicted wage is the linear projection of the log of wages on explanatory variables, including sex, age, the presence of spouse, the number of children, the number of pre-unemployment contributions during the last 3 years, tenure and its square, and a set of fixed effects for eligibility group, month and year.

\begin{figure}[!h]
\caption{Predicted wages}\label{f:covariates2}
\begin{center}
\begin{center}
\subfigure[2005 to March 2006]{\includegraphics*[width=7.5cm,height=5cm]{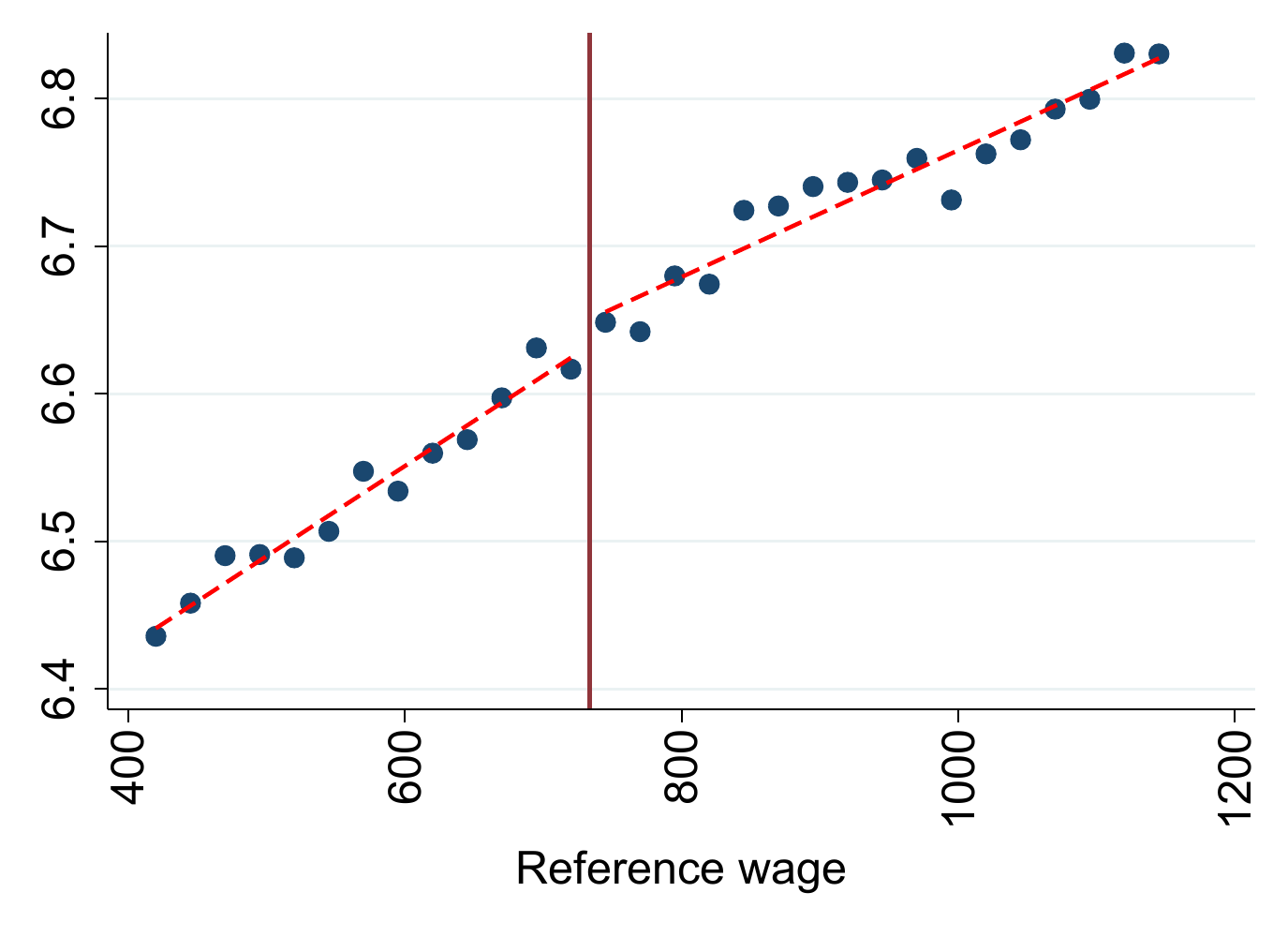}}
\subfigure[April 2006 to 2009]{\includegraphics*[width=7.5cm,height=5cm]{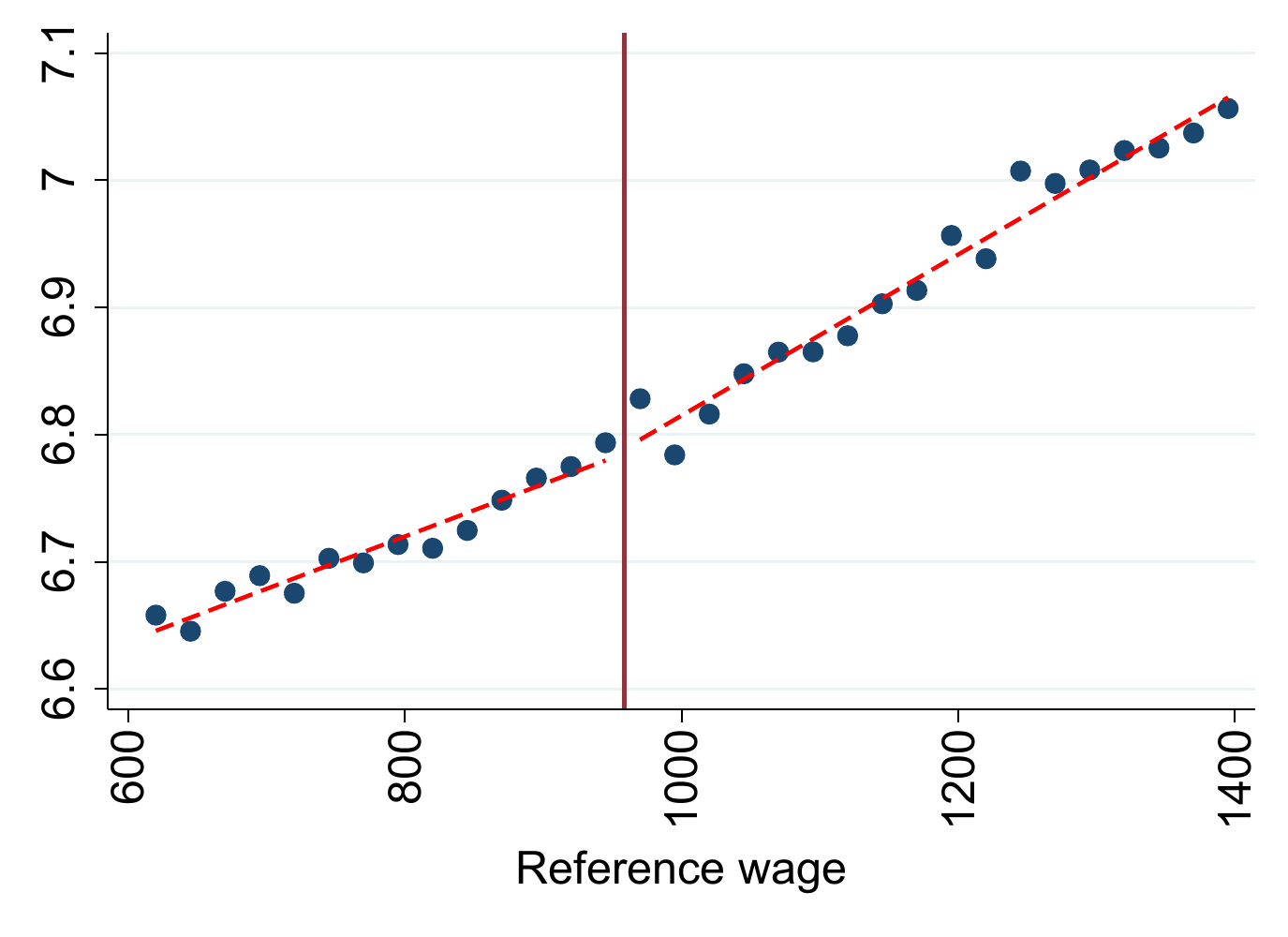}}
\end{center}
\begin{minipage}{0.9\textwidth} \footnotesize
Note: The predicted wages is the result of a linear projection of the log of re-employment wages on a set of covariates, including the reference wage, age, sex, presence of spouse, number of children, and the number of periods with contributions in the previous three years. See notes to Figure \ref{f:Btot}.
\end{minipage}
\end{center}
\end{figure}

\begin{figure}[!h]
\caption{Predicted total benefits paid}\label{f:covariates}
\begin{center}
\begin{center}
\subfigure[2005 to March 2006]{\includegraphics*[width=7.5cm,height=5cm]{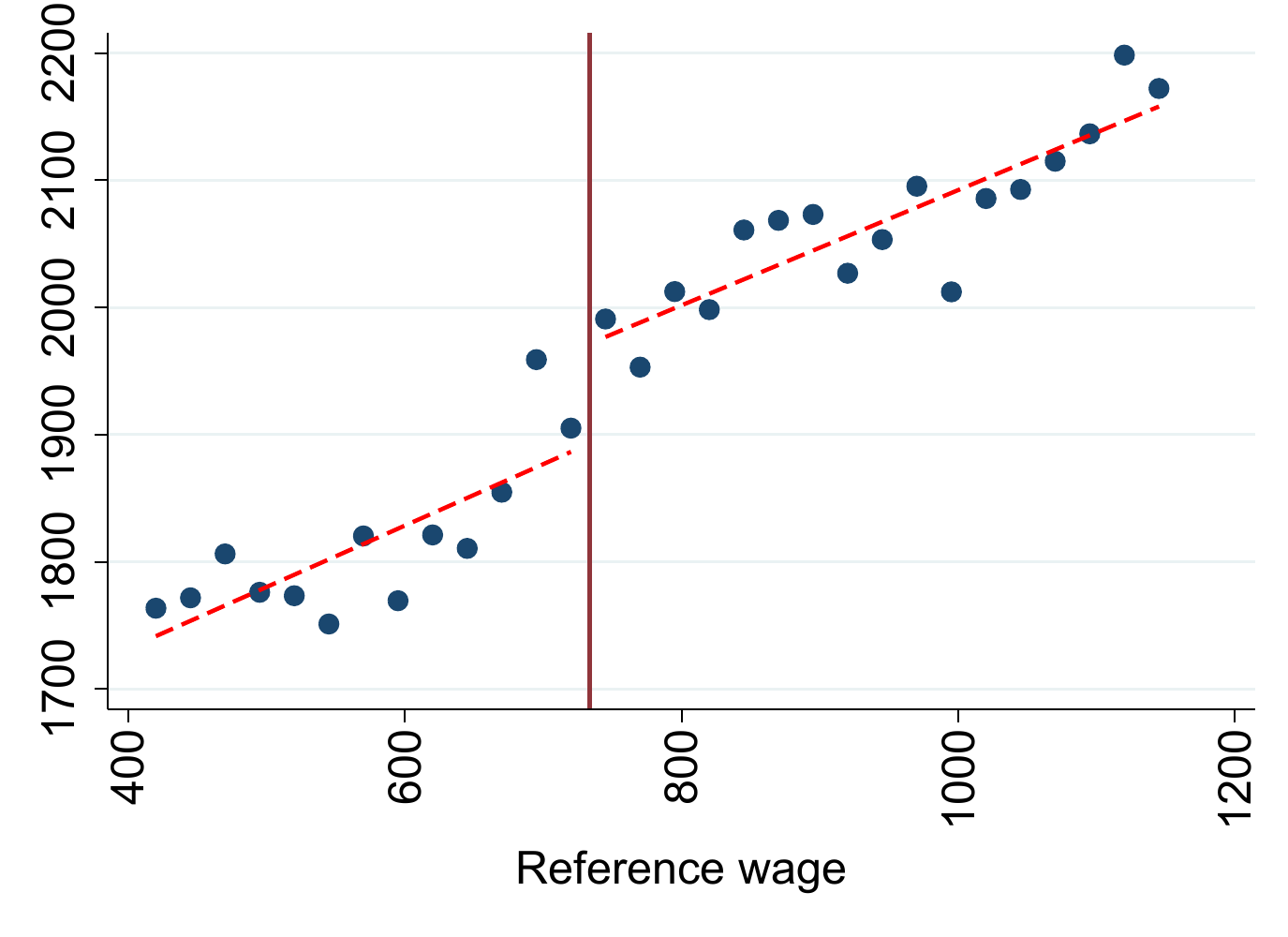}}
\subfigure[April 2006 to 2009]{\includegraphics*[width=7.5cm,height=5cm]{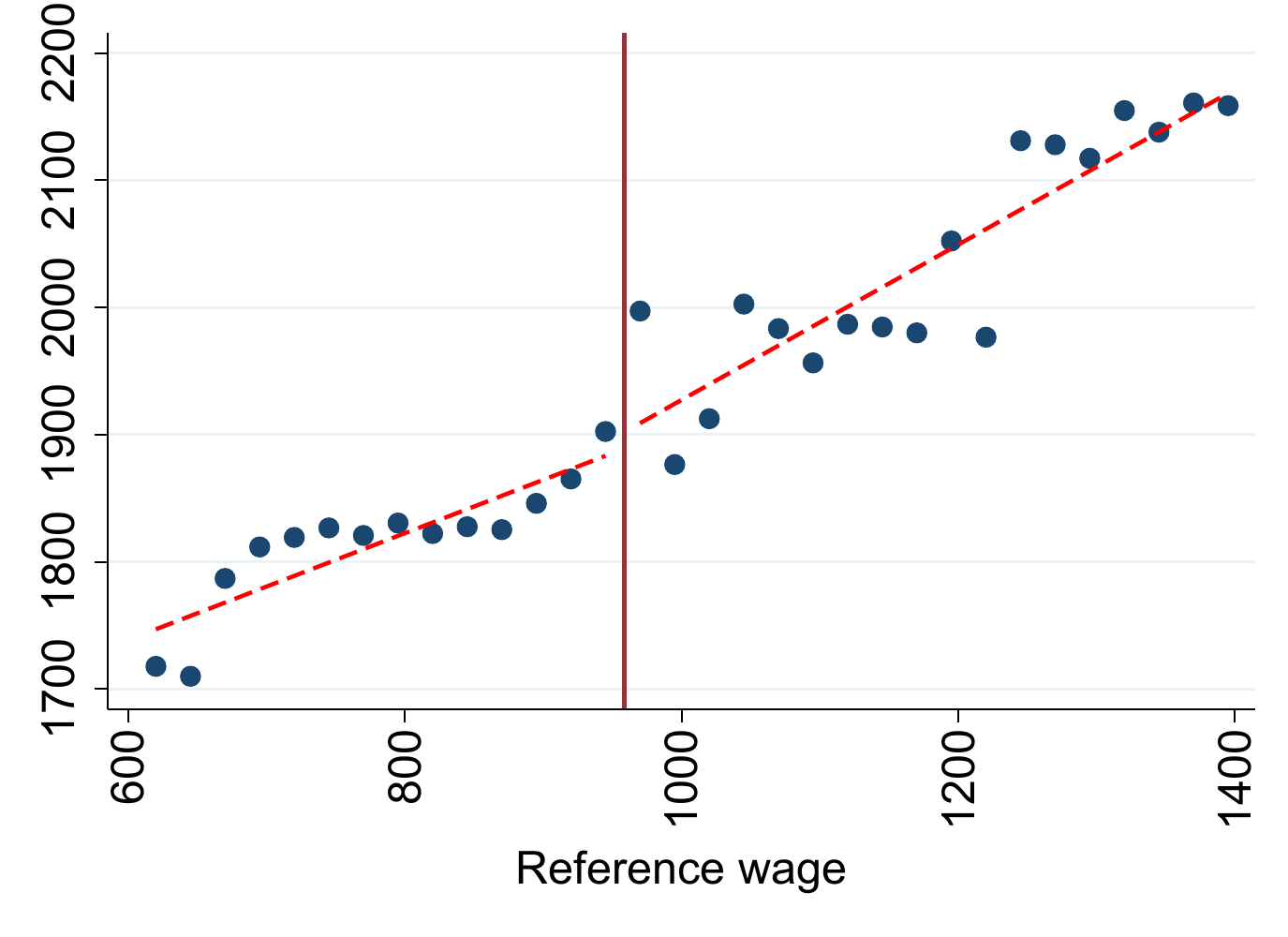}}
\end{center}
\begin{minipage}{0.9\textwidth} \footnotesize
Note: The predicted total benefits paid is the result of a linear projection of that variable on a set of covariates, including the reference wage, age, sex, presence of spouse, number of children, and the number of periods with contributions in the previous three years. See notes to Figure \ref{f:Btot}.
\end{minipage}
\end{center}
\end{figure}

\figref{f:covariates2} plots the predicted value of log wages on the assignment variable. There is no appreciable change in the slopes of the predicted re-employment wages at the kinks. This suggests that the kinks that arise in \figref{f:lnwage} arise in the outcome variable, it is generated not by discontinuities in the covariates at both sides of the kink but by the kink in the benefit level. Figure \ref{f:covariates} reproduce the same results for the total UI paid.

\begin{figure}[!h]
\caption{Distribution of the assignment variable}\label{f:mccrary}
\begin{center}
\subfigure[2005 to March 2006]{\includegraphics*[width=7.5cm,height=5cm]{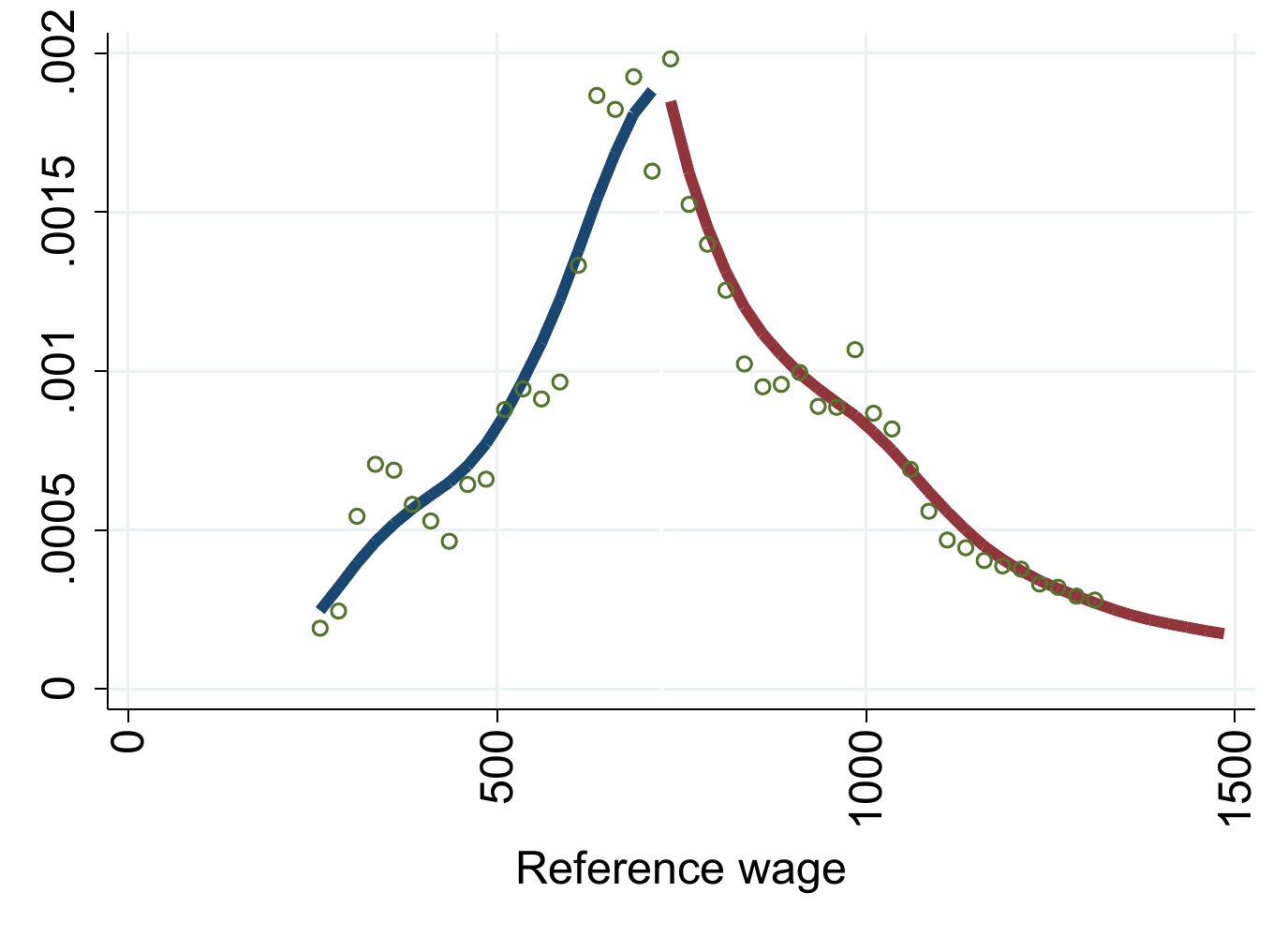}}
\subfigure[April 2006 to 2009]{\includegraphics*[width=7.5cm,height=5cm]{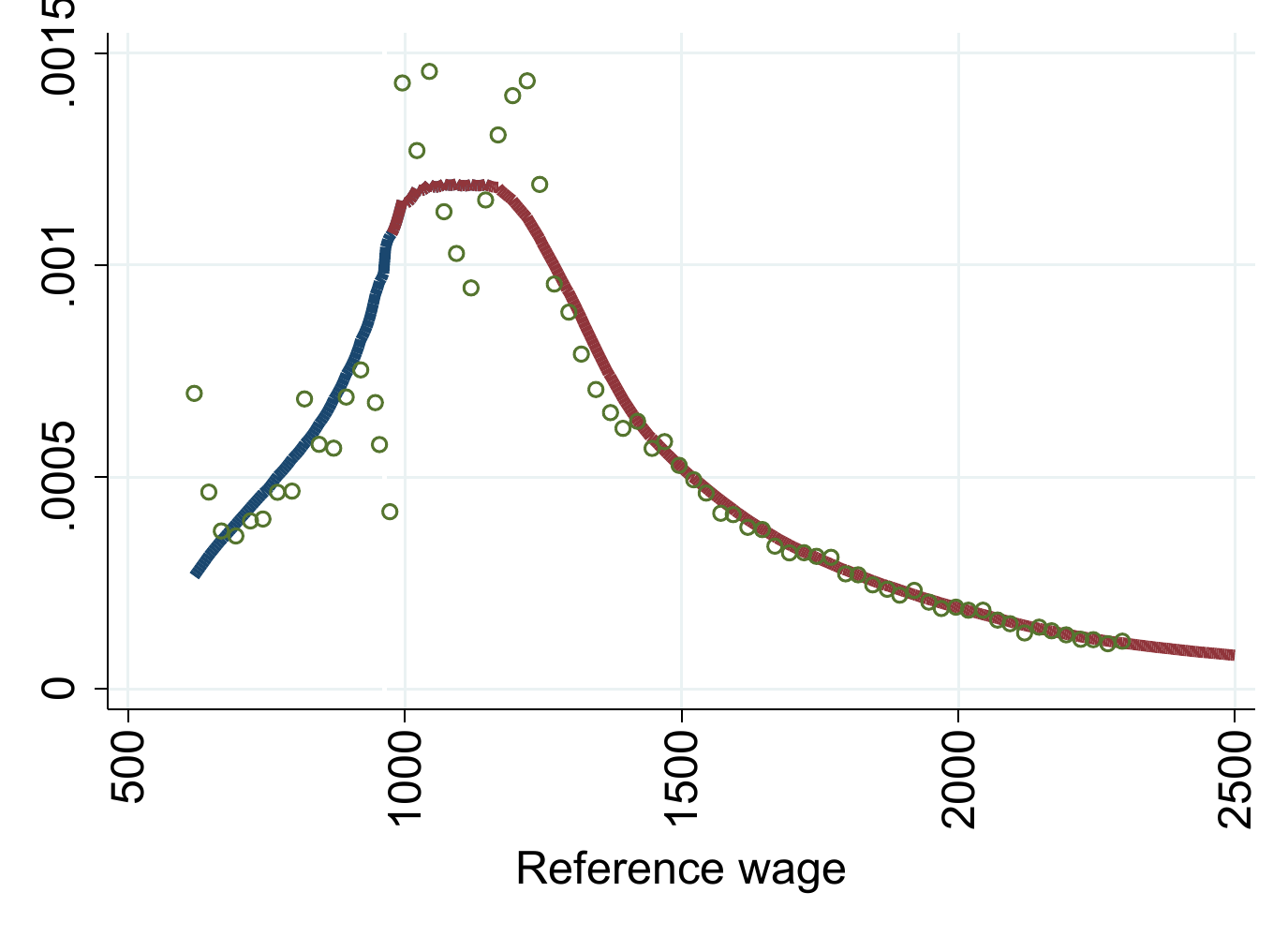}}
\end{center}
\begin{center}
\begin{minipage}{0.9\textwidth} \footnotesize
Note: The graph shows the number of observations within bins of the assignment variable (the gross reference wage).
\end{minipage}
\end{center}
\end{figure}

Finally, we turn to consider the issue of the manipulation of the assignment variable. To address this issue, \figref{f:mccrary} plots the number of observed spells of UI as a function of the assignment variable and analyzes the discontinuity at the top kink. From the panel (a) it can be seen that the top kink is close to the mode of the distribution and that there is no jump in the relationship between the number of observations and the assignment variable at the kink point. In panel (b) it is also apparent that the mode has moved to the right of the kink. Again, there is no jump in the density at kink point.
Importantly, RKD methods assume that both the density and its partial derivative are continuous at the kink. The plot in panel (b) satisfy this assumption, given that there is no jump and also no change in slope at the kink. In panel (a) there is continuity but a possible change in slope. Nevertheless, we interpret this as exclusively related to the fact that the kink is at the mode and not because of a manipulation of running variable or the policy.

\subsection{Estimation results}\label{S:Estimation}

We now turn to analyze the estimation results. Our purpose is to report the effects of benefit level on the re-employment wages and on the total amount of benefits paid. We do so by estimating equation (\ref{eq:frkd}) using IV and applying an FG bandwidth. We concentrate on the data at the high kink after the reform. We postpone to Section \ref{S:Robustness} the robustness of our results to samples, bandwidths, methods and quadratic specifications.

\tabref{tab:frkd} reports the results of estimating the effects of UI on total UI paid and on re-employment wages. For the two variables we show results using no controls in columns (1) and (3) and a set of controls in columns (2)\ and (4).

As anticipated by \figref{f:schedule0} the first-stage regression coefficient is always significant and very precisely estimated. All coefficients
are -0.41, indicating that the slope changes from 0.41 to zero at $k_1^H$. Notice that we  observe gross pre-unemployment reference wage while the actual benefit is computed as the 50\% of the net reference wage. In any case, the first stage shows that the IV is highly relevant.

The  IV coefficient is positive and significant, meaning that both re-employment wage and total UI paid increase with benefits. In particular, we estimate that the log of re-employment wages increase by about 0.001. This result can be transformed in an elasticity by multiplying the coefficient by the value of benefits at the kink to get $\frac{d \ln w}{d b} b^{1}_{H}$. This elasticity is 0.37 and changes to 0.23 when we add controls.

The IV coefficient for the total amount of benefits paid is of 6.5 with no controls and 6.2 with controls. Considering that around the kink the average level of paid UI is 2585 ARS, we compute an elasticity of 1.01. These results suggest that the possible behavioral effect of the change in effective replacement rate should be very low, suggesting that the changes in UI benefits does not strongly affect the job-finding probability and the amount of eligible periods.

\subsection{Calibration of the formula}\label{S:Calibration}

We now go back to the main formula in equation (\ref{eq:FOCf}) and use our estimates to calibrate it.
We rewrite our formula as
\begin{equation}\label{eq:welfare}
\frac{\partial {\overline{\log w}} }{\partial b} \times b = \frac{\frac{\partial R}{\partial b}} {\frac{1}{\delta}+R/b}
\end{equation}
where we provide an underestimation of welfare gains by setting $w^{\ast}=b$ and $\frac{\sigma^2}{\sigma^2-v_i}=1$. Notice that the choice of setting the reservation wage to the level of benefits tends to reduce the left hand side of the equation. It is reasonable to expect the reservation wage to be higher than the benefit paid. First, the replacement rate is below 50\% at the kinks used for identification, and we can expect that the average re-employment wage to be close to the pre-unemployment wage.
Second, the observed re-employment wage is almost always above the amount of the benefit; in particular, about 90\% of workers get a re-employment wage higher than the initial benefit level. Notice that the welfare gains are also underestimated when using the lower bound of $\frac{\sigma^2}{\sigma^2-var}$ which is 1.

The left hand side of equation (\ref{eq:FOC1}) measures the welfare gain through the change in the reservation wage of the eligible workers. In practice, we use all re-employment wages to gain observations. According to a theoretical search model, the effect of any benefit on re-employment wages would be lower after exhaustion. Thus, by considering all re-employment wages, we are understating the effect. For these reasons, our basic calibration of the left hand side of the formula can be considered as a lower bound of the welfare gains of increasing UI level.

Both left hand side (welfare gains) and right hand side (welfare losses)\ can be measured through the results presented above. The only exception is that we need an estimation of $1/\delta$. That parameter represents the average duration in employment or average tenure measured at layoff. From the data we have that this value goes from around 33 to 47 (see \tabref{tbl:characteristics}). We set then $\delta=0.03$ as the high end of the separation rate or the low end of the distribution in tenure. Again, our purpose is to report a lower bound of the net welfare gains.\footnote{The separation rate used is based on the pre-unemployment average employment duration; re-employment duration is probably higher, given that separation rate is lower as workers age.}

\tabref{tab:welfare} reports the left hand side and right hand side of the welfare formula, considering both periods and with and without controls. The right hand side is 0.16 and the left hand side ranges from 0.23 to 0.37. In all cases, the point estimate of the net welfare gains, the left hand side minus the right hand side of the formula, are positive. Additionally, we base on coefficients that are highly significant.

From this calibration of the formula we conclude that there are substantial net welfare gains in rising the benefits for the unemployed. In particular, welfare gains are equivalent to 21 cents for each additional monetary unit of benefit.
Taking into account that these are underestimations of the actual welfare gains according to the model, the result remarkably points out the convenience of increasing UI\ benefit in Argentina above the levels considered in our sample.

Overall, these results and conclusions tend to be in line with the results in \citeN{GRRuffo2016}. That paper estimated relatively low social costs and high wage elasticities for the increase in UI level. In this paper we show similar conclusions exploiting a very different source of identification.

\section{Robustness}\label{S:Robustness}

We now turn to explore how results change when we consider different samples and alternative specifications of the RKD method.
We provide robustness in a variety of dimensions. We first maintain the method and provide results for different samples. We then vary methods and bandwidths to show the sensitivity of the estimates to these changes. Finally, we provide alternative specifications using the samples before and after the reform.

\subsection{Pre-reform results}
We estimate the same RKD methods using UI spells with exhausted eligibility  before the reform. In this case, we use the observations around $k^0_H$.  

\tabref{tab:frkd0} reports the results of estimating the effects of UI benefits on wages and total UI paid in the spell. The implied elasticities of wages are very similar to the ones after the reform. The effects on the total UI paid are even lower than in the previous case. All results are highly significant, even when the sample is much smaller than the one of our main results.

\tabref{tab:frkd0}  also presents the results of calibrating the main formula with these estimates and provides a very similar conclusion: there are welfare gains of increasing UI.  The welfare gains using point estimates of column (1) (without controls) suggest a 21 cents welfare gain for each additional monetary unit increase in benefits, the same as our main result. When using the results of column (2), with controls, we find a 14 cents increase in welfare.

\subsection{Subsamples}
The effects of UI are not necessarily homogeneous for different workers. Differences in re-employment opportunities, in expected re-employment wages, in the time to retirement, in the duration of jobs, all may contribute to heterogeneous search behavior and UI effects by, for example, age, gender, industry and region. In our particular UI design, the potential duration of UI change according to the age of the worker and the number of contributions in the last 36 months (see \tabref{tbl:tab2}). In what follows we divide the sample according to these two variables.

\tabref{tab:samples1}
shows the results for workers younger than 45 years of age, column (1), and workers of 45 years or older, column (3). We find that the elasticities for young workers are close to the aggregate. In particular, the elasticity of re-employment wages is around 0.30. The effects on the total amount of benefits paid were of 6.6, in line with the aggregate effects. The results for older workers differ more: their wage elasticities is higher (0.9) and the effect on the total UI paid are larger (10.8). This comes at no surprise since the potential duration and expected UI duration are longer for this group. In all, the evaluation of the welfare formulas, using these highly significant results, give a welfare net gain of 0.15 for the young and 0.67 for those older than 45 years of age.

Columns (2) and (4) of \tabref{tab:samples1}
present the analogous estimations for the subsample composed by workers with pre-unemployment contributions between 24 and 35 months. Under this additional restriction, potential duration is homogeneous in 8 months for workers younger than 45 years of age and in 14 months for workers with 45 years of age or older. We choose this group because it is the one that more closely relates to the average potential duration in the sample, which is between   9 to 10 months. Our conclusions of positive welfare gains hold for these two groups. We acknowledge that the results for older workers are less stable, perhaps affected by the limited sample size in that group.

\subsection{Bandwidths and specifications}

RKD results are sensitive to the the particular implementation, polynomial order and bandwidth \cite{cardetal2012}. For that reason, we present results for different bandwidths, for linear and quadratic specifications, and for sharp and fuzzy methods.

\tabref{tab:bandwiths}
presents the results for a sharp-RKD estimation using a fixed bandwidth of 100, in column (1), for the optimal (MSRERD) bandwidth with triangular weights\footnote{See \citeN{Calonicoetal}.}, in column (2), for the linear fuzzy RKD design with a fixed bandwidth at 200 without additional controls, in column (3), and with controls, in column (4), a quadratic specification for the fuzzy RKD, with FG bandwidth, in column (5), with a fixed bandwidth at 200, in column (6), and with controls, in column (7). As anticipated, the elasticities of re-employment wages vary substantially. Nevertheless, linear specifications with sufficiently wide bandwidths provide elasticities between 0.17 and 0.7. The elasticities of the benefits paid are more stable, ranging from 1.04 and 1.32. Excluding the exceptions with very narrow bandwidths, we find a positive point estimation of the welfare gains, many of them based on highly significant estimates.

\subsection{Alternative identification strategies using several kinks}

In our baseline specification we analyze the data after the reform of 2006, and we show that results do not change much when we consider the pre-reform sample. In this section we use the data both before and after the reform.
Consider the following equations,
\begin{eqnarray}
Y &=&\mu _{0}+\alpha b+\gamma _{1}\widetilde{w}+X^{\prime }\beta _{1}+\mu
_{1T}T+u  \label{eq:strp} \\
b &=&\delta _{0}+\delta _{1}\widetilde{w}+\delta _{2}\widetilde{w}\times
D+\mu _{2T}T+X^{\prime }\beta _{2}+v  \label{eq:fsp}
\end{eqnarray}
where the endogenous variable $b$ in (\ref{eq:strp}) is instrumented in equation (\ref{eq:fsp}) by $\widetilde{w}\times D$, where $D =1\left\{ \widetilde{w}\geq 0\right\}$. In  these equations $T$ is a dummy variable identifying all periods after the reform, and
$$\widetilde{w} = \left( w-k_{H}^0\right) \left( 1-T\right) +\left(
w-k_{H}^1\right) T$$
is a transformation of the running variable, the reference wage, that normalizes its value to zero at the kink of the corresponding period. This implementation is a direct extension of the RKD method.\footnote{The 2006 reform also allows for the estimation of a RKD\ in Double-Difference as in \citeN{Landais15}. Nevertheless, the fact that the change was relatively small impose the use of small bandwidth and limited sample size for which results are not robust.}

\tabref{tab:bothtop} reports the estimates using a variety of methods and bandwidths using both samples, before and after the reform.
Column (1) reports sharp linear RKD with a small bandwidth (100 ARS). The effect on wages is positive, implying an elasticity of 0.57, but the result is not significant. The effect on total UI\ paid is highly significant but low, implying an elasticity of 0.85. Column (2) reports the result of using an optimal bandwidth, weighting observations with a triangular
kernel. Results are similar to the ones in the first column.

Columns (3), (4) and (5) also report similar results from a linear model. Column (3) applies an FG bandwidth without controls, while column (5) adds controls. Column (4) implements a much smaller bandwidth of 200.
Columns (6) to (8) report the result of a quadratic specification for different bandwidths. Results vary more in this case. For example, for an FG bandwidth without controls the wage elasticity is 0.97 while the effect on the UI paid implies an elasticity of 1.05;\ these results change for a bandwidth of 200 ARS.

The bottom panel of the table reports the computed welfare gains. In all cases, welfare gains are positive. For linear specifications results range from 0.24 to 0.43; for the quadratic specifications the results for welfare gains are much larger.

\section{Conclusions}\label{S:Conclusions}

This paper estimated alternative indicators of the effects of UI benefit level. We exploited individual kinks in the schedule of initial UI transfers implementing regression kink design methods. We find, on the one hand, an elasticity close to one on the total amount paid of UI per spell of covered unemployment. This means that the number of covered UI months only changes modestly, so that behavioral effects are small. On the other hand, we find a significant effect on re-employment wages. This suggests a better outcome in the search process. Jointly, these two findings imply that a more generous UI level could be convenient for workers.

To formalize this assessment, we use a sufficient statistic approach and evaluate the welfare gains and social costs of an increase in UI level using these estimates. By calibrating an optimality formula, we find that any increase in UI level of one monetary unit generates a welfare gain of between 7 and 21 cents. Given that our measures are underestimating actual welfare gains, this result is remarkable.

To interpret our results it is important to keep in mind that the maximum benefits represented between 20\% and 25\% of the average gross wages in the private formal sector in Argentina for the periods analyzed.  In this sense, our conclusion that the level of benefits should be increased comes at no surprise given that the initial level is relatively low. Of course that at the kink the net replacement rate is about 40\%, but the incentives of UI is not related to past wages but to the re-employment wages, which are probably higher than pre-unemployment wages.

Importantly, the RKD methods that we use identify a weighted average marginal effect at the kink, where the weights reflect the relative likelihood of being at the kink for each individual. Importantly, the kinks used for identification are close to the median of the distribution of the assignment variable, and the characteristics of workers at the kink are similar to the average characteristics of the population. Thus, we can consider these results as relevant to evaluate the effects of UI level on the median beneficiary.

At the same time, the method compares individuals with higher or lower replacement rates in an otherwise similar environment. Macroeconomic and labor market conditions are the same for workers at both sides of the kink. Any endogenous effects that could be related to the increase in benefits, such as tax rise, prices adjustments, or even aggregate labor supply reductions, are kept unchanged. In this sense, these methods identify partial equilibrium effects, and are more easily interpretable than other empirical implementations, such as those exploiting changes in benefits between regions or wide reforms.

Finally, our results are very relevant to understand the effects of government transfers in an economy with high informality. We show that the increase of the level of UI  does not generate a strong disincentive. Thus, a transfer program can be welfare improving even with the additional moral hazard of informality.


\clearpage
\bibliographystyle{chicago}

\begin{thebibliography}{}

\bibitem[\protect\citeauthoryear{Bover, Arellano, and Bentolila}{Bover
  et~al.}{2002}]{ArellanoBoverBentolila}
Bover, O., M.~Arellano, and S.~Bentolila (2002).
\newblock Unemployment duration, benefit duration and the business cycle.
\newblock {\em Economic Journal\/}~{\em 112\/}(479), 223--265.

\bibitem[\protect\citeauthoryear{Calonico, Cattaneo, and Titiunik}{Calonico
  et~al.}{2014}]{Calonicoetal}
Calonico, S., M.~D. Cattaneo, and R.~Titiunik (2014).
\newblock {Robust Nonparametric Confidence Intervals for
  Regression-Discontinuity Designs}.
\newblock {\em Econometrica\/}~{\em 82}, 2295--2326.


\bibitem[\protect\citeauthoryear{Campos, Garc\'{i}a-P\'{e}rez, and
  Reggio}{Campos et~al.}{2017}]{camposetal}
Campos, R.~G., J.~I. Garc\'{i}a-P\'{e}rez, and I.~Reggio (2017).
\newblock {Moral Hazard versus Liquidity and the Optimal Timing of Unemployment
  Benefits}.
\newblock Working Papers 2017-10, FEDEA.

\bibitem[\protect\citeauthoryear{Card, Chetty, and Weber}{Card
  et~al.}{2007}]{CardChettyWeber}
Card, D., R.~Chetty, and A.~Weber (2007).
\newblock Cash-on-hand and competing models of intertemporal behavior: New
  evidence from the labor market.
\newblock {\em Quarterly Journal of Economics\/}~{\em 122\/}(4), 1511--1560.

\bibitem[\protect\citeauthoryear{Card, Lee, Pei, and Weber}{Card
  et~al.}{2012}]{cardetal2012}
Card, D., D.~Lee, Z.~Pei, and A.~Weber (2012).
\newblock {Nonlinear Policy Rules and the Identification and Estimation of
  Causal Effects in a Generalized Regression Kink Design}.
\newblock NBER Working Papers 18564, National Bureau of Economic Research, Inc.

\bibitem[\protect\citeauthoryear{Card, Lee, Pei, and Weber}{Card
  et~al.}{2015}]{Cardetal2015}
Card, D., D.~S. Lee, Z.~Pei, and A.~Weber (2015).
\newblock {Inference on Causal Effects in a Generalized Regression Kink
  Design}.
\newblock {\em Econometrica\/}~{\em 83}, 2453--2483.

\bibitem[\protect\citeauthoryear{Centeno and Novo}{Centeno, M. and A. Novo}{2009}]{CentenoNovo09}
Centeno, M. and A. Novo (2009).
\newblock {Reemployment wages and UI liquidity effect: a regression
  discontinuity approach}.
\newblock {\em Portuguese Economic Journal\/}~{\em 8\/}(1), 45--52.


\bibitem[\protect\citeauthoryear{Chetty}{Chetty}{2008}]{Chetty08}
Chetty, R. (2008).
\newblock {Moral Hazard versus Liquidity and Optimal Unemployment Insurance}.
\newblock {\em Journal of Political Economy\/}~{\em 116\/}(2), 173--234.

\bibitem[\protect\citeauthoryear{Chetty}{Chetty}{2009}]{Chetty09SuffStats}
Chetty, R. (2009).
\newblock {Sufficient Statistics for Welfare Analysis: A Bridge Between
  Structural and Reduced-Form Methods}.
\newblock {\em Annual Review of Economics\/}~{\em 1\/}(1), 451--488.

\bibitem[\protect\citeauthoryear{Chetty and Finkelstein}{Chetty and
  Finkelstein}{2012}]{Chetty-Finkelstein}
Chetty, R. and A.~Finkelstein (2012).
\newblock Social insurance: Connecting theory to data.
\newblock NBER Working Papers 18433, National Bureau of Economic Research, Inc.

\bibitem[\protect\citeauthoryear{Dong}{Dong}{2011}]{Dong2011}
Dong, Y. (2011).
\newblock {Jumpy or Kinky? Regression Discontinuity without the Discontinuity}.
\newblock Working Papers 111207, University of California-Irvine, Department of
  Economics.

\bibitem[\protect\citeauthoryear{Fan and Gijbels}{Fan and Gijbels}{1996}]{FG}
Fan, J. and I.~Gijbels (1996).
\newblock {\em {Local Polynomial Modelling and Its Applications}}.
\newblock Chapman and Hall.

\bibitem[\protect\citeauthoryear{Gerard and Gonzaga}{Gerard and
  Gonzaga}{2016}]{gerard-gonzaga}
Gerard, F. and G.~Gonzaga (2016).
\newblock {Informal Labor and the Efficiency Cost of Social Programs: Evidence
  from the Brazilian Unemployment Insurance Program}.
\newblock NBER Working Papers 22608, National Bureau of Economic Research, Inc.

\bibitem[\protect\citeauthoryear{Gonzalez-Rozada and Ruffo}{Gonzalez-Rozada and
  Ruffo}{2016}]{GRRuffo2016}
Gonzalez-Rozada, M. and H.~Ruffo (2016).
\newblock {Optimal unemployment benefits in the presence of informal labor
  markets}.
\newblock {\em Labour Economics\/}~{\em 41\/}(C), 204--227.

\bibitem[\protect\citeauthoryear{Gonzalez-Rozada, Ronconi, and
  Ruffo}{Gonzalez-Rozada et~al.}{2011}]{GRRR}
Gonzalez-Rozada, M., L.~Ronconi, and H.~Ruffo (2011).
\newblock Protecting workers against unemployment in latin america and the
  caribbean: Evidence from argentina.
\newblock Technical report, IDB Working Paper No. IDB-WP-268.

\bibitem[\protect\citeauthoryear{Katz and Meyer}{Katz and
  Meyer}{1990}]{Katzmeyer90}
Katz, L.~F. and B.~D. Meyer (1990).
\newblock The impact of the potential duration of unemployment benefits on the
  duration of unemployment.
\newblock {\em Journal of Public Economics\/}~{\em 41\/}(1), 45--72.

\bibitem[\protect\citeauthoryear{Lalive}{Lalive}{2007}]{Lalive07AER}
Lalive, R. (2007).
\newblock {Unemployment Benefits, Unemployment Duration, and Post-Unemployment
  Jobs: A Regression Discontinuity Approach}.
\newblock {\em American Economic Review\/}~{\em 97\/}(2), 108--112.

\bibitem[\protect\citeauthoryear{Landais}{Landais}{2015}]{Landais15}
Landais, C. (2015).
\newblock {Assessing the Welfare Effects of Unemployment Benefits Using the
  Regression Kink Design}.
\newblock {\em American Economic Journal: Economic Policy\/}~{\em 7\/}(4),
  243--278.

\bibitem[\protect\citeauthoryear{Landais, Michaillat, and Saez}{Landais
  et~al.}{2018}]{LandaisMichaillatSaez18ap}
Landais, C., P.~Michaillat, and E.~Saez (2018).
\newblock {A Macroeconomic Approach to Optimal Unemployment Insurance:
  Applications}.
\newblock {\em American Economic Journal: Economic Policy\/}~{\em 10\/}(2),
  182--216.

\bibitem[\protect\citeauthoryear{Lee, Leung, O'Leary, Pei, and Quach}{Lee
  et~al.}{2021}]{Leeetal2021}
Lee, D.~S., P.~Leung, C.~J. O'Leary, Z.~Pei, and S.~Quach (2021).
\newblock {Are Sufficient Statistics Necessary? Nonparametric Measurement of
  Deadweight Loss from Unemployment Insurance}.
\newblock {\em Journal of Labor Economics\/}~{\em 39\/}(S2), 455--506.



\bibitem[\protect\citeauthoryear{Machin and Manning}{Machin and
  Manning}{1999}]{Machin99}
Machin, S. and A.~Manning (1999).
\newblock Chapter 47: The causes and consequences of longterm unemployment in
  europe.
\newblock Volume 3, Part 3 of {\em Handbook of Labor Economics}, pp.\  3085 --
  3139. Elsevier.

\bibitem[\protect\citeauthoryear{McCall}{McCall}{1970}]{Mccall70}
McCall, J. (1970).
\newblock Economics of information and job search.
\newblock {\em Quarterly Journal of Economics\/}~{\em 84\/}(1), 113--126.

\bibitem[\protect\citeauthoryear{Meyer}{Meyer}{1990}]{meyer90}
Meyer, B. (1990).
\newblock Unemployment insurance and unemployment spells.
\newblock {\em Econometrica\/}~{\em 58\/}(4).

\bibitem[\protect\citeauthoryear{Nekoei and Weber}{Nekoei and
  Weber}{2017}]{Nekoei-Weber}
Nekoei, A. and A.~Weber (2017).
\newblock {Does Extending Unemployment Benefits Improve Job Quality?}
\newblock {\em American Economic Review\/}~{\em 107\/}(2), 527--561.

\bibitem[\protect\citeauthoryear{Nielsen, S{\o}rensen, and Taber}{Nielsen
  et~al.}{2010}]{Nielsenetal}
Nielsen, H.~S., T.~S{\o}rensen, and C.~Taber (2010).
\newblock {Estimating the Effect of Student Aid on College Enrollment: Evidence
  from a Government Grant Policy Reform}.
\newblock {\em American Economic Journal: Economic Policy\/}~{\em 2\/}(2),
  185--215.

\bibitem[\protect\citeauthoryear{Schmieder, von Wachter, and Bender}{Schmieder
  et~al.}{2012}]{Schmiederetal12}
Schmieder, J.~F., T.~von Wachter, and S.~Bender (2012).
\newblock {The Effects of Extended Unemployment Insurance Over the Business
  Cycle: Evidence from Regression Discontinuity Estimates Over 20 Years}.
\newblock {\em The Quarterly Journal of Economics\/}~{\em 127\/}(2), 701--752.

\bibitem[\protect\citeauthoryear{Schmieder, von Wachter, and Bender}{Schmieder
  et~al.}{2016}]{Schmieder-vonWachter-Bender16AER}
Schmieder, J.~F., T.~von Wachter, and S.~Bender (2016).
\newblock {The Effect of Unemployment Benefits and Nonemployment Durations on
  Wages}.
\newblock {\em American Economic Review\/}~{\em 106\/}(3), 739--777.

\bibitem[\protect\citeauthoryear{Shimer and Werning}{Shimer and
  Werning}{2007}]{ShimerWerning07}
Shimer, R. and I.~Werning (2007).
\newblock Reservation wages and unemployment insurance.
\newblock {\em Quarterly Journal of Economics\/}~{\em 122\/}(3), 1145--1185.

\end{thebibliography}

\clearpage

\section{Tables}

\begin{table}[h]
\begin{center}%
\caption{Unemployment Insurance Eligibility}%
\label{tbl:tab2}
\vspace{0.25cm}
{\normalsize
\begin{tabular}{ccc}
\hline
\hline
Months with & \multicolumn{2}{c}{Months of UI support}  \\
contributions to UI & if age $<$ 45 & if age $\geq 45$ \\
during the last & & \\
36 months &  &  \\ \hline
6 to 11 & 2 & 8 \\
12 to 23 & 4 & 10 \\
24 to 35 & 8 & 14 \\
36 or more & 12 & 18 \\ \hline \hline
\end{tabular}
}

\vspace{0.25cm}
\begin{minipage}{1\textwidth}
\footnotesize
Note: Before March 2006, only workers with at least 12 months of contributions
during the last 36 months were eligible. The contributions can be both continuous or discontinuous. The system of temporary contracts and of construction sector are not considered within this paper.
\end{minipage}
\end{center}
\end{table}%

\begin{table}[tbp]
\caption{Descriptive statistics by period and range of reference wage} \label{tbl:characteristics}
\vspace{0.25cm}
{\small
\begin{tabular}{l | c | ccc | ccc}
\hline
\hline
&(1)&(2)&(3)&(4)&(5)&(6)&(7) \\
&&&&&&& \\
\hline
 & Total & \multicolumn{3}{c}{Jan.2005 to Mar.2006} & \multicolumn{3}{c}{Apr.2006 to Dec.2007} \\
Minimum reference wage& 75 & 75&361&722&     75& 602&963 \\
Maximum reference wage&4800& 361&722&3600&  602& 963&4800 \\
\midrule
Reference wage&1186.45&306.70&596.82&1115.04&494.32&811.35&1525.36 \\
UI initial ben.&338.84&150.10&244.13&299.60&250.13&332.99&399.88 \\
UI potential duration&9.45&9.09&9.43&10.56&8.66&9.08&10.29 \\
\midrule Age&35.43&34.54&34.35&36.02&34.68&34.48&35.37 \\
Male&0.69&0.49&0.65&0.75&0.45&0.60&0.73 \\
Dependent spouse&0.51&0.45&0.49&0.56&0.44&0.48&0.55 \\
Dependent children&0.77&0.78&0.85&0.88&0.67&0.73&0.80 \\
Tenure in pre-un. job&38.86&35.28&38.05&46.96&32.99&35.45&45.57 \\
Number of contrib.&28.09&27.14&28.17&30.44&26.33&27.47&30.15 \\
\midrule Non-emp. dur.&21.13&31.54&27.04&23.12&24.20&21.48&18.70 \\
UI duration&7.09&7.06&7.20&7.54&7.14&7.22&7.55 \\
Total UI transfers&2213.38&1194.80&1690.87&2178.32&1785.66&2192.74&2663.79 \\
Num.of obs&114914&2687&16565&25369&9406&17792&70662 \\
\hline
\hline
\end{tabular}%
}
\vspace{0.25cm}
\begin{minipage}{0.99\textwidth}
\footnotesize
Notes: The table divide the sample according to the initial period of UI and the reference wage. The reported reference wage is the highest gross wage of the last six. Bottom kink is at 361 ARS before March 2006 and 602 afterwards; top kink is at 722 ARS before March 2006 and 963 afterwards. Columns (2) and (5) include workers below the bottom kink; columns (4) and (7) workers above the top kink; columns (3) and (6) include workers in between the kinks. The schedule of initial benefits depends on net reference wage, while we observe gross reference wage.
\end{minipage}
\end{table}%

\clearpage

\begin{table}[htbp]
  \centering
  \caption{Estimates of the effect of benefits using the high kink}\label{tab:frkd}%
    \vspace{0.25cm}
\begin{tabular}{lrlrlrlrrlrl}
\hline
\hline
           &                \multicolumn{ 5}{c}{Log of re-employment wages} &            &                             \multicolumn{ 5}{c}{Total UI paid} \\

           & \multicolumn{ 2}{c}{(1)} &            & \multicolumn{ 2}{c}{(2)} &            & \multicolumn{ 2}{c}{(3)} &            & \multicolumn{ 2}{c}{(4)} \\

           & \multicolumn{ 2}{c}{No controls} &            & \multicolumn{ 2}{c}{Controls} &            & \multicolumn{ 2}{c}{No controls} &            & \multicolumn{ 2}{c}{Controls} \\
\hline
\textit{April 2006 to 2009} &            &            &            &            &            &            &            &            &            &            &            \\
\\

$\pi_1$, first stage coef. &    -0.4147 &        *** &            &    -0.4145 &        *** &            &    -0.4143 &        *** &            &    -0.4141 &        *** \\

           &   (0.0003) &            &            &   (0.0005) &            &            &   (0.0003) &            &            &   (0.0004) &            \\

$\alpha^{RKD}$, RKD coef. &     0.0009 &        *** &            &     0.0006 &         ** &            &       6.53 &        *** &            &       6.20 &        *** \\

           &   (0.0002) &            &            &   (0.0002) &            &            &   (0.1785) &            &            &     (0.12) &            \\


$\overline{Y}$, mean of outcome &       7.05 &            &            &       7.05 &            &            &    2585.57 &            &            &    2585.57 &            \\

$\eta$, elasticity &       0.37 &            &            &       0.23 &            &            &       1.01 &            &            &       0.96 &            \\

$h$, bandwidth &    1657.02 &            &            &    1657.02 &            &            &    1624.98 &            &            &    1624.98 &            \\

         N &      48994 &            &            &      44291 &            &            &      82064 &            &            &      77062 &            \\
\hline
\hline
\end{tabular}

      \vspace{0.25cm}
    \begin{minipage}{0.99\textwidth}
    \footnotesize
    Note: The coefficients are the result of implementing IV methods to estimate the effect of benefit level. It reports fuzzy RKD estimates as in equation (\ref{eq:frkd}) using the periods when the maximum level of benefits were at 400 ARS.
Robust standard errors for the estimates are in parentheses. The elasticities are computed as in equation (\ref{eq:elasticity}), where $b_{k}$ is the initial benefit at the kink and $\bar{Y}$ is the average of the variable at the kink, reported in the table. Controls include age, age squared, gender, number of children, presence of spouse, pre-unemployment tenure and its square, number of pre-unemployment contributions, imputed severance pay level, and identification variables for the eligibility UI duration, region, industry of pre-unemployment job, year, and month; for the log of re-employment wages, column (2), controls include a set of dummies identifying groups of observations by four months of non-employment duration (less than 5, 5 to 8, etc.).  Statistical significance: * significant at the 10\% level; ** significant at the 5\% level; *** significant at the 1\% level.
    \end{minipage}
\end{table}%

\clearpage

\begin{table}[htbp]
  \centering
  \caption{Calibration of the welfare formula}\label{tab:welfare}%
    \vspace{0.25cm}
\begin{tabular}{lrr}
\hline
\hline
        &       No controls             &       Controls                \\

\hline
\textit{April 2006 to 2009}    &            &            \\
\\
LHS: $\eta_{w,b}$       &       0.37    ***     &       0.23    **      \\
        &                       &                       \\
RHS:    &       0.16    ***     &       0.16    ***     \\
$R/b$   &       6.46               &       6.46     \\
$\delta$        &       0.03            &       0.03            \\
        &                       &                       \\
Welfare gains (LHS-RHS) &       0.21            &       0.07            \\

\hline
\hline
\end{tabular}

      \vspace{0.25cm}
    \begin{minipage}{0.75\textwidth}
    \footnotesize
    Note: The table reports the calibration of equation (\ref{eq:welfare}) with the estimates of Table \ref{tab:frkd}.
    \end{minipage}
\end{table}%

\clearpage

\begin{table}[htbp]
  \centering
  \caption{Estimates of the effect of benefits using the high kink before the reform}\label{tab:frkd0}%
    \vspace{0.25cm}
\begin{tabular}{lrrrrr}

           &            &            &            &            &            \\
\hline
\hline
           & \multicolumn{ 2}{c}{(1)} &            & \multicolumn{ 2}{c}{(2)} \\

           & \multicolumn{ 2}{c}{No controls} &            & \multicolumn{ 2}{c}{Controls} \\
\hline
           &            &            &            &            &            \\

                      \multicolumn{ 6}{l}{{\it A. Log of re-employment wages}} \\

$\alpha^{RKD}$, RKD coef.  &     0.0012 &        *** &            &     0.0009 &         ** \\

           &   (0.0003) &            &            &   (0.0004) &            \\

           &            &            &            &            &            \\

$\overline{Y}$, mean of outcome  &       6.79 &            &            &       6.79 &            \\

$\eta$, elasticity &       0.36 &            &            &       0.27 &            \\

$h$, bandwidth &    4289.62 &            &            &    4289.62 &            \\

         N &      12496 &            &            &      11321 &            \\

           &            &            &            &            &            \\
\hline
           &            &            &            &            &            \\

                                   \multicolumn{ 6}{l}{{\it B. Total UI paid}} \\

$\alpha^{RKD}$, RKD coef. &       5.69 &        *** &            &       5.08 &        *** \\

           &   (0.1846) &            &            &     (0.14) &            \\

           &            &            &            &            &            \\

$\overline{Y}$, mean of outcome &    1995.95 &            &            &    1995.95 &            \\

$\eta$, elasticity &       0.86 &            &            &       0.76 &            \\

$h$, bandwidth &    1288.37 &            &            &    1288.37 &            \\

         N &      17145 &            &            &      15949 &            \\

           &            &            &            &            &            \\
\hline
           &            &            &            &            &            \\

\textit{C. Welfare formula}           &            &            &            &            &            \\

LHS: $\eta_{w,b}$ &       0.36 &            &            &       0.27 &            \\

           &            &            &            &            &            \\

       RHS &       0.14 &            &            &       0.13 &            \\

    $R/b$ &       6.65 &            &            &       6.65 &            \\

 $\delta$ &       0.03 &            &            &       0.03 &            \\

Welfare gains &       0.21 &            &            &       0.14 &            \\

\hline
\hline
           &            &            &            &            &            \\

\end{tabular}

      \vspace{0.25cm}
    \begin{minipage}{0.99\textwidth}
    \footnotesize
    Note: The coefficients are the result of implementing IV methods to estimate the effect of benefit level on the pre-reform sample, when the maximum level of benefits were at 300 ARS (see notes to Table \ref{tab:frkd}). Robust standard errors for the estimates are in parentheses. Statistical significance: * significant at the 10\% level; ** significant at the 5\% level; *** significant at the 1\% level.
    \end{minipage}
\end{table}%

\clearpage


\begin{table}[htbp]
  \centering
  \caption{Estimations different samples}\label{tab:samples1}%
    \vspace{0.25cm}
\begin{tabular}{lrlrlrlrl}
\hline \hline

           & \multicolumn{ 2}{c}{(1)} & \multicolumn{ 2}{c}{(2)} & \multicolumn{ 2}{c}{(3)} & \multicolumn{ 2}{c}{(4)} \\

           & \multicolumn{ 4}{c}{Age $<$ 45} & \multicolumn{ 4}{c}{Age $\geq$ 45}  \\
           & \multicolumn{ 2}{c}{all} & \multicolumn{ 2}{c}{contr. 24 to 35} & \multicolumn{ 2}{c}{all} & \multicolumn{ 2}{c}{contr. 24 to 35} \\
\hline
\multicolumn{ 3}{l}{Top Kink - April 2006 to 2009 }            &            &            &            &            &            &            \\

\multicolumn{ 3}{l}{\it A. Log of re-employment wages}               &            &            &            &            &            &            \\

$\alpha^{RKD}$, RKD coef &     0.0008 &        *** &     0.0007 &         ** &     0.0023 &        *** &     0.0010 &            \\

           &   (0.0002) &            &   (0.0004) &     {\bf } &   (0.0007) &            &   (0.0008) &            \\

           &            &            &     {\bf } &     {\bf } &            &            &            &            \\

$\overline{Y}$, mean of outcome &       7.03 &            &       7.01 &     {\bf } &       7.17 &            &       7.19 &            \\

$\eta$, elasticity &       0.32 &            &       0.30 &     {\bf } &       0.92 &            &       0.40 &            \\

$h$, bandwidth &    1699.60 &            &    2629.33 &     {\bf } &    2952.84 &            &    5304.01 &            \\

         N &      43106 &            &      12739 &     {\bf } &       6541 &            &       1535 &            \\

           &            &            &            &            &            &            &            &            \\
\hline
           &            &            &            &            &            &            &            &            \\

{\it B. Total UI paid} &            &            &     {\bf } &     {\bf } &     {\bf } &     {\bf } &     {\bf } &     {\bf } \\

$\alpha^{RKD}$, RKD coef &       6.61 &        *** &       5.78 &        *** &      10.76 &        *** &       9.36 &        *** \\

           &    (0.156) &            &   (0.2245) &     {\bf } &    (0.468) &     {\bf } &   (0.8491) &     {\bf } \\

 &            &            &     {\bf } &     {\bf } &            &     {\bf } &     {\bf } &     {\bf } \\

$\overline{Y}$, mean of outcome &    2269.37 &            &    2232.46 &     {\bf } &    3916.90 &     {\bf } &    3561.69 &     {\bf } \\

$\eta$, elasticity &       1.17 &            &       1.04 &     {\bf } &       1.10 &     {\bf } &       1.05 &     {\bf } \\

$h$, bandwidth &    1849.76 &            &    2355.42 &     {\bf } &    2321.61 &     {\bf } &    2719.04 &     {\bf } \\

         N &      67867 &            &      19025 &     {\bf } &      15816 &     {\bf } &       3214 &     {\bf } \\

           &            &            &     {\bf } &     {\bf } &     {\bf } &     {\bf } &     {\bf } &     {\bf } \\

           &            &            &     {\bf } &     {\bf } &     {\bf } &     {\bf } &     {\bf } &     {\bf } \\
\hline
           &            &            &     {\bf } &     {\bf } &     {\bf } &     {\bf } &     {\bf } &     {\bf } \\
\textit{C. Welfare formula}           &            &            &            &            &            \\

LHS: $\eta_{w,b}$ &       0.32 &            &       0.30 &     {\bf } &       0.92 &     {\bf } &       0.40 &     {\bf } \\

       RHS &       0.17 &            &       0.15 &     {\bf } &       0.25 &     {\bf } &       0.22 &     {\bf } \\

    $R/b$ &       5.67 &            &       5.58 &     {\bf } &       9.79 &     {\bf } &       8.90 &     {\bf } \\

 $\delta$ &       0.03 &            &       0.03 &            &       0.03 &            &       0.03 &            \\

Welfare gains &       0.15 &            &       0.15 &            &       0.67 &            &       0.18 &            \\

           &            &            &            &            &            &            &            &            \\
\hline
\hline
\end{tabular}
      \vspace{0.25cm}
    \begin{minipage}{0.95\textwidth}
    \footnotesize
    Note: The coefficients are the result of implementing IV methods to estimate the effect of benefit level. It reports fuzzy RKD estimates by groups of workers defined by age and number of pre-unemployment contributions.  Robust standard errors for the estimates are in parentheses. See notes to Table \ref{tab:frkd}. Statistical significance: * significant at the 10\% level; ** significant at the 5\% level; *** significant at the 1\% level..
    \end{minipage}
\end{table}%


\begin{landscape}

\begin{table}[htbp]
  \centering
  \caption{Estimations using different bandwdiths}\label{tab:bandwiths}%
    \vspace{0.25cm}
    \footnotesize
\begin{tabular}{lrlrlrlrlrlrlrl}
\hline \hline
           & \multicolumn{ 2}{c}{(1)} & \multicolumn{ 2}{c}{(2)} & \multicolumn{ 2}{c}{(3)} & \multicolumn{ 2}{c}{(4)} & \multicolumn{ 2}{c}{(5)} & \multicolumn{ 2}{c}{(6)} & \multicolumn{ 2}{c}{(7)} \\

           & \multicolumn{ 2}{c}{Sharp} & \multicolumn{ 2}{c}{Linear} & \multicolumn{ 2}{c}{Linear } & \multicolumn{ 2}{c}{Linear} & \multicolumn{ 2}{c}{Quadratic} & \multicolumn{ 2}{c}{Quadratic} & \multicolumn{ 2}{c}{Quadratic} \\
          & \multicolumn{ 2}{c}{     } & \multicolumn{ 2}{c}{Opt MSERD} & \multicolumn{ 2}{c}{  } & \multicolumn{ 2}{c}{  } & \multicolumn{ 2}{c}{FG} & \multicolumn{ 2}{c}{ } & \multicolumn{ 2}{c}{ } \\
\hline
\multicolumn{ 3}{l}{Top Kink - April 2006 to 2009}             &            &            &            &            &            &            &            &            &            &            &            &            \\

\multicolumn{ 3}{l}{{\it A. Log of re-employment wages}}            &            &            &            &            &            &            &            &            &            &            &            &            \\

$\alpha^{RKD}$ &    -0.0002 &            &    -0.0007 &            &     0.0004 &        *** &     0.0009 &            &     0.0025 &        *** &     0.0014 &            &     0.0033 &            \\

           &   (0.0017) &            &   (0.0014) &            &   (0.0006) &            &   (0.0007) &            &   (0.0007) &            &   (0.0025) &            &   (0.0026) &            \\

           &            &            &            &            &            &            &            &            &            &            &            &            &            &            \\

Controls           & \multicolumn{ 2}{c}{NO} & \multicolumn{ 2}{c}{NO} & \multicolumn{ 2}{c}{NO} & \multicolumn{ 2}{c}{YES} & \multicolumn{ 2}{c}{NO} & \multicolumn{ 2}{c}{NO} & \multicolumn{ 2}{c}{YES} \\

$\overline{Y}$ &       7.05 &            &       7.05 &            &       7.05 &            &       7.05 &            &       7.05 &            &       7.05 &            &       7.05 &            \\

$\eta$ &      -0.09 &            &      -0.27 &            &       0.17 &            &       0.35 &            &       0.99 &            &       0.55 &            &       1.33 &            \\

$h$ &     100.00 &            &     131.08 &            &     200.00 &            &     200.00 &            &    2349.02 &            &     200.00 &            &     200.00 &            \\

         N &      10760 &            &      13448 &            &      19282 &            &      17557 &            &      50631 &            &      19282 &            &      17557 &            \\

           &            &            &            &            &            &            &            &            &            &            &            &            &            &            \\
\hline
           &            &            &            &            &            &            &            &            &            &            &            &            &            &            \\

\multicolumn{ 3}{l}{{\it B. Total UI paid}}            &            &            &            &            &            &            &            &            &            &            &            &            \\

$\alpha^{RKD}$ &       6.70 &        *** &       7.97 &        *** &       6.85 &        *** &       6.71 &        *** &       8.27 &        *** &       8.15 &        *** &       8.50 &        *** \\

           &   (1.8033) &            &   (1.9965) &            &   (0.6452) &            &   (0.4113) &            &   (0.6321) &            &   (2.5367) &            &   (1.5549) &            \\

 &            &            &            &            &            &            &            &            &            &            &            &            &            &            \\

Controls           & \multicolumn{ 2}{c}{NO} & \multicolumn{ 2}{c}{NO} & \multicolumn{ 2}{c}{NO} & \multicolumn{ 2}{c}{YES} & \multicolumn{ 2}{c}{NO} & \multicolumn{ 2}{c}{NO} & \multicolumn{ 2}{c}{YES} \\
$\overline{Y}$ &    2585.57 &            &    2585.57 &            &    2585.57 &            &    2585.57 &            &    2585.57 &            &    2585.57 &            &    2585.57 &            \\

$\eta$ &       1.04 &            &       1.23 &            &       1.06 &            &       1.04 &            &       1.28 &            &       1.26 &            &       1.32 &            \\

$h$ &     100.00 &            &     109.13 &            &     200.00 &            &     200.00 &            &    1865.11 &            &     200.00 &            &     200.00 &            \\

         N &      18016 &            &      19328 &            &      32332 &            &      30484 &            &      83288 &            &      32332 &            &      30484 &            \\

           &            &            &            &            &            &            &            &            &            &            &            &            &            &            \\

\hline
\multicolumn{ 3}{l}{{\it C. Welfare formula}}                 &            &            &            &            &            &            &            &            &            &            &            &            \\

           &            &            &            &            &            &            &            &            &            &            &            &            &            &            \\

LHS: $\eta_{w,b}$ &      -0.09 &            &      -0.27 &            &       0.17 &            &       0.35 &            &       0.99 &            &       0.55 &            &       1.33 &            \\

       RHS &       0.17 &            &       0.20 &            &       0.17 &            &       0.17 &            &       0.21 &            &       0.20 &            &       0.21 &            \\

    $R/b$ &       6.46 &            &       6.46 &            &       6.46 &            &       6.46 &            &       6.46 &            &       6.46 &            &       6.46 &            \\

 $\delta$ &       0.03 &            &       0.03 &            &       0.03 &            &       0.03 &            &       0.03 &            &       0.03 &            &       0.03 &            \\

Welfare gains &      -0.26 &            &      -0.47 &            &       0.00 &            &       0.19 &            &       0.79 &            &       0.34 &            &       1.11 &            \\

           &            &            &            &            &            &            &            &            &            &            &            &            &            &            \\
\hline
\hline
\end{tabular}
      \vspace{0.25cm}
    \begin{minipage}{0.99\textwidth}
    \footnotesize
    Note: The coefficients are the result of implementing IV methods to estimate the effect of benefit level for the post-reform sample, when the maximum level of benefits were at 400 ARS. Columns present different specifications and bandwidths. Robust standard errors for the estimates are in parentheses. See notes to Table \ref{tab:frkd}. Statistical significance: * significant at the 10\% level; ** significant at the 5\% level; *** significant at the 1\% level.
    \end{minipage}
\end{table}%

\clearpage

\begin{table}[htbp]
  \centering
  \caption{Estimations using high kink in both pre and post-reform }\label{tab:bothtop}%
    \vspace{0.25cm}
    \scriptsize
\begin{tabular}{lrlrlrlrlrlrlrlrl}
\hline
\hline
           & \multicolumn{ 2}{c}{(1)} & \multicolumn{ 2}{c}{(2)} & \multicolumn{ 2}{c}{(3)} & \multicolumn{ 2}{c}{(4)} & \multicolumn{ 2}{c}{(5)} & \multicolumn{ 2}{c}{(6)} & \multicolumn{ 2}{c}{(7)} & \multicolumn{ 2}{c}{(8)} \\

           & \multicolumn{ 2}{c}{Sharp} & \multicolumn{ 2}{c}{Linear}   & \multicolumn{ 2}{c}{Linear} & \multicolumn{ 2}{c}{Linear } & \multicolumn{ 2}{c}{Linear } & \multicolumn{ 2}{c}{Quadratic } & \multicolumn{ 2}{c}{Quadratic } & \multicolumn{ 2}{c}{Quadratic } \\

           & \multicolumn{ 2}{c}{} & \multicolumn{ 2}{c}{opt.$\dagger$} & \multicolumn{ 2}{c}{ FG} & \multicolumn{ 2}{c}{200  } &   \multicolumn{ 2}{c}{200} & \multicolumn{ 2}{c}{ FG} & \multicolumn{ 2}{c}{ 200} & \multicolumn{ 2}{c}{ 200 } \\
\hline
           &            &            &            &            &            &            &            &            &            &            &            &            &            &            &            &            \\

         \multicolumn{ 5}{l}{{\it A. Log of re-employment wages}} &            &            &            &            &            &            &            &            &            &            &            &            \\

$\alpha^{RKD}$ &     0.0014 &            &     0.0012 &            &     0.0010 &        *** &     0.0011 &         ** &     0.0010 &          * &     0.0024 &        *** &     0.0035 &          * &     0.0044 &         ** \\

           &   (0.0013) &            &   (0.0012) &            &   (0.0002) &            &   (0.0005) &            &   (0.0005) &            &   (0.0006) &            &   (0.0019) &            &    (0.002) &            \\

           &            &            &            &            &            &            &            &            &            &            &            &            &            &            &            &            \\

  Controls &         NO &            &         NO &            &         NO &            &         NO &            &        YES &            &         NO &            &         NO &            &        YES &            \\

$\overline{Y}$ &       6.96 &            &       6.96 &            &       6.96 &            &       6.96 &            &       6.96 &            &       6.96 &            &       6.96 &            &       6.96 &            \\

$\eta$ &       0.57 &            &       0.48 &            &       0.40 &            &       0.43 &            &       0.39 &            &       0.97 &            &       1.42 &            &       1.77 &            \\

$h$ &     100.00 &            &     120.51 &            &    1342.95 &            &     200.00 &            &     200.00 &            &    2053.25 &            &     200.00 &            &     200.00 &            \\

         N &      15923 &            &      18438 &            &      59833 &            &      27373 &            &      24906 &            &      62537 &            &      27373 &            &      24906 &            \\

           &            &            &            &            &            &            &            &            &            &            &            &            &            &            &            &            \\
\hline
           &            &            &            &            &            &            &            &            &            &            &            &            &            &            &            &            \\

                      \multicolumn{ 5}{l}{{\it B. Total UI paid}} &            &            &            &            &            &            &            &            &            &            &            &            \\

$\alpha^{RKD}$ &       5.50 &        *** &       6.07 &        *** &       5.93 &        *** &       6.94 &        *** &       5.99 &        *** &       6.81 &        *** &       5.71 &        *** &       6.82 &        *** \\

           &   (1.3435) &            &    (1.191) &            &   (0.1544) &            &    (0.508) &            &    (0.338) &            &   (0.5012) &            &   (1.9346) &            &   (1.2436) &            \\

           &            &            &            &            &            &            &            &            &            &            &            &            &            &            &            &            \\

  Controls &         NO &            &         NO &            &         NO &            &         NO &            &        YES &            &         NO &            &         NO &            &        YES &            \\

$\overline{Y}$ &    2585.57 &            &    2585.57 &            &    2585.57 &            &    2585.57 &            &    2585.57 &            &    2585.57 &            &    2585.57 &            &    2585.57 &            \\

$\eta$ &       0.85 &            &       0.94 &            &       0.92 &            &       1.07 &            &       0.93 &            &       1.05 &            &       0.88 &            &       1.05 &            \\

$h$ &     100.00 &            &     129.58 &            &    1040.83 &            &     200.00 &            &     200.00 &            &    1510.80 &            &     200.00 &            &     200.00 &            \\

         N &      25377 &            &      31129 &            &      93248 &            &      43758 &            &      41134 &            &      98558 &            &      43758 &            &      41134 &            \\

           &            &            &            &            &            &            &            &            &            &            &            &            &            &            &            &            \\
\hline
           &            &            &            &            &            &            &            &            &            &            &            &            &            &            &            &            \\

                      \multicolumn{ 5}{l}{{\it C. Welfare formula}} &            &            &            &            &            &            &            &            &            &            &            &            \\
LHS: $\eta_{w,b}$ &       0.57 &            &       0.48 &            &       0.40 &            &       0.43 &            &       0.39 &            &       0.97 &            &       1.42 &            &       1.77 &            \\

       RHS &       0.14 &            &       0.15 &            &       0.15 &            &       0.17 &            &       0.15 &            &       0.17 &            &       0.14 &            &       0.17 &            \\

    $R/b$ &       6.46 &            &       6.46 &            &       6.46 &            &       6.46 &            &       6.46 &            &       6.46 &            &       6.46 &            &       6.46 &            \\

 $\delta$ &       0.03 &            &       0.03 &            &       0.03 &            &       0.03 &            &       0.03 &            &       0.03 &            &       0.03 &            &       0.03 &            \\

Welfare gains &       0.43 &            &       0.33 &            &       0.25 &            &       0.25 &            &       0.24 &            &       0.80 &            &       1.27 &            &       1.60 &            \\

           &            &            &            &            &            &            &            &            &            &            &            &            &            &            &            &            \\
\hline
\hline
\end{tabular}        \vspace{0.25cm}
    \begin{minipage}{0.99\textwidth}
    \scriptsize
    Note: The coefficients are the result of implementing IV methods to estimate the effect of benefit level for both pre- and post-reform samples. Columns present different specifications and bandwidths. Robust standard errors for the estimates are in parentheses. See notes to Table \ref{tab:frkd}. Statistical significance: * significant at the 10\% level; ** significant at the 5\% level; *** significant at the 1\% level. \\
$\dagger$ Optimal MSERD bandwidth and triangular kernel.
    \end{minipage}
\end{table}%

\end{landscape}

\clearpage

\clearpage

\section{Figures}

\clearpage

\clearpage


\appendix
\section*{Appendix \label{S:Appendix}}
\renewcommand\thefigure{\thesection.\arabic{figure}}
\setcounter{figure}{0}
\renewcommand\thetable{\thesection.\arabic{table}}
\setcounter{table}{0}
\renewcommand\theequation{\thesection.\arabic{equation}}
\setcounter{equation}{0}

\section{Model}\label{S:AppModel}

Consider a representative worker that derives consumption utility from a concave  function, $u(c)$, and discounts the future at rate $r$. While employed the worker receives net income $w-\tau$, where $w$ is the wage at the job and $\tau$ is a labor income tax. There is a constant probability $\delta$ of being laid off from the job, in which case, the worker becomes unemployed. The expected lifetime utility of the worker in a job with wage $w$ satisfies \begin{equation}\label{eq:apmodelV} rV(w) = u(w-\tau) + \delta \left[ U-V(w) \right],\end{equation}

\noindent where $U$ represents the expected lifetime utility of an eligible unemployed worker. After displacement, the worker becomes eligible for receiving unemployment insurance, $b$, each period. The worker loses eligibility to UI\ at a constant probability $\gamma$. Besides the benefit, the worker collects an (informal, ie. unobserved by the government) income $y$ if not employed. While searching for a new job, the worker receives a wage offer with probability $\lambda$. Wage offers are a random draw from the $F(w)$ distribution function.\ Under these conditions, the lifetime utility of an eligible worker satisfies $$rU=u(b+y)+ \lambda \int \max \{ V(w)-U,0 \}dF(w) + \gamma \left[ S-U \right],$$
where $S$ is the lifetime utility of an unemployed worker after losing UI\ eligibility,
$$rS=u(b_{a}+y)+\lambda  \int \max \{ V(w)-S,0 \}dF(w).$$
where $b_a$ is an unlimited assistance transfer. Under these conditions, the worker would set a constant reservation wage, $w^{\ast}$ while eligible unemployed, and a different (lower)\ reservation wage after losing eligibility. The reservation wage after displacement would be the solution to $rV(w^{\ast})=rU$. At the same time, from equation (\ref{eq:apmodelV}), $rV(w^{\ast})=u(w^{\ast}-\tau)$. Thus, maximizing the lifetime utility of an eligible unemployed worker after displacement would be equivalent to maximize the after tax reservation wage, $w^{\ast}-\tau$. The main intuition is that, in this setup, the net reservation wage is the monetary counterpart of the utility of the eligible unemployed \cite{ShimerWerning07}.

Consider an increase in the level of benefits. The first order condition is

$$\frac{\partial w^{\ast}}{\partial b} + \frac{\partial w^{\ast}}{\partial \tau} \tau'- \tau'=0,$$
where $\tau'$ is the total effect of benefits on taxes. To further simplify this formula, we follow Shimer and Werning by assuming CARA preferences, by which we can get $\frac{\partial w^{\ast}}{\partial b} = \frac{\partial w^{\ast}}{\partial \tau}$, so that the welfare formula can be written as in equation (\ref{eq:FOC1}).
Thus, an increase in benefits would improve the welfare of the eligible unemployed whenever $\frac{\partial w^{\ast}}{\partial b}>\frac{\tau'}{1+\tau'} $.

The planner would maximize the net reservation wage subject to the incentive compatibility constraint and the budget constraint. We will consider a simple per-period budget constraint, by which all income from employed workers should equate the expenditure on UI benefits.\footnote{Let $\tau e=b u$, where $e$ is the proportion of employed workers, and $u$ is the proportion of eligible unemployed workers. The inflows and outflows to $u$ should balance in steady state so that $\delta e = (\gamma + \lambda \left[1-F(w^{\ast})\right])u$. This means that we can write the budget constraint as $b \delta\ = \tau (\gamma + \lambda \left[1-F(w^{\ast})\right])$. Given that the average duration in states is the inverse of the (constant) rate out of the  state, the budget constraint can be stated in terms of expected duration.
} Then, the budget constraint can be written  $$bB=D_E\tau$$
where $B=1/(\gamma + \lambda \left[1-F(w^{\ast})\right])$ is the expected duration in UI, and $D_E=1/\delta$ is the expected duration in a job. To capture the effects of benefits on the budget, we consider that any change in benefits and in taxes would affect the reservation wage and, thus, the expected duration $B$. In particular,
$$B db + \frac{\partial B}{\partial b} db + \frac{\partial B}{\partial \tau} d\tau \ = \frac{ d\tau}{\delta} $$
so that
$$ \frac{\tau' }{1+\tau'}= \frac{B+ \frac{\partial B}{\partial b} }{( 1/\delta -  \frac{\partial B}{\partial \tau} + \frac{\partial B}{\partial b}  +B)}  $$
where $\tau' = d\tau /db$ is the total change in taxes after a change in benefits, including the behavioral responses of the same tax changes. Under CARA preferences $\frac{\partial w^{\ast}}{\partial b} = \frac{\partial w^{\ast}}{\partial \tau}$, which also means $\frac{\partial B}{\partial b} = \frac{\partial B}{\partial \tau}$, and defining the total amount of UI paid as $R=bB$ so that $\partial R/ \partial b=B+\partial B/\partial b$, we can write the above equation as $$ \frac{\tau' }{1+\tau'}= \frac{\partial R/\partial b}{1/ \delta+R/b}  $$

In this way we get a simple expression for the RHS of the welfare formula: it depends crucially on the response of the total UI paid to the UI level. This response has two components: the mechanical effect ($B$) and the behavioral effect ($\partial B/\partial b$). Our method in this paper is simply to measure the total effect $\partial R/\partial b$. 

The LHS\ of the formula shows that the net reservation wage of the eligible unemployed worker is our sufficient statistic for the welfare gains. While reservation wages are unobservable, we can infer the changes in the reservation wage from changes in re-employment wages. For that purpose, we assume that workers receive wage offers from a log normal distribution, $\log(w) \sim \mathcal{N}(\mu,\sigma)$. The distribution of accepted wages would be, then, truncated by the reservation wage, so that

$$\overline{\log( w)}=\int_{\log(w^{\ast})}^\infty \log (w) \frac{f(\log(w))}{1-F( \log( w^\ast))}dw. $$

\noindent This implies that  $\frac{\partial \overline{\log( w)} }{\partial b} = \frac{\partial \log(w^{\ast}) }{\partial b} \frac{f( \log(w^{\ast}))}{[1-F(\log(w^{\ast}))]}[\overline{ \log(w)}-\log(w^{\ast})]$. At the same time, the variance of the log accepted wages would satisfy $var=Var[\log(w)|w\geq w^{\ast}]= \sigma^2-\sigma \frac{ \phi(\frac{\log(w^{\ast})-\mu}{\sigma})}{1-\Phi(\frac{\log(w^{\ast})-\mu}{\sigma})} [\overline{\log(w)}-\log(w^{\ast})]$, where $\phi()$ and $\Phi()$ are the pdf and the cdf of the  standard normal distribution. Given that in this case $ \frac{1}{\sigma} \times \frac{ \phi(\frac{\log(w^{\ast})-\mu}{\sigma})}{1-\Phi(\frac{\log(w^{\ast})-\mu}{\sigma})}=\frac{f( \log(w^{\ast}))}{[1-F(\log(w^{\ast}))]}$ we can approximate the change in the reservation wage by the following formula: $$ \frac{\partial w^{\ast} }{\partial b} =  \frac{\partial \overline{\log( w)}}{\partial b}  w^{\ast} \Lambda.$$
where $\Lambda=\frac{\sigma^2}{\sigma^2-var}$.
Notice that we now use the change in the average log re-employment wage  as the sufficient statistic.
The problem is now to give reasonable values to the remaining parameters of the formula.
For that purpose we assume that the reservation wage is close to the benefit level, which is a conservative assumption, and we impose $\Lambda=1$, the low end of the value of that parameter. This implies that the LHS\ of the formula subsumes to the elasticity of accepted wages to benefits, $\eta_{w,b}$. Now, we finally get

$$ \eta_{w,b}\ =  \frac{ \partial R/\partial b}{1/ \delta+R/b} $$

The formula compares the response of re-employment wages with the response in total UI paid. Importantly, this formula does not depend on the informal income, $y$, in a direct way. Informality can, in fact, affect the optimal level of UI, but its effects would be through the values of parameters and elasticities of the formula.

While the presentation uses a particular simple setup based on hand-to-mouth agents, the formula is robust to a number of extensions.  \citeN{ShimerWerning07} show that the idea that the effect of UI can be captured through the change in reservation wages arises when workers can save, or can exert search effort. It is also  valid when unemployment benefits are provided for limited periods, and when jobs differ in separation rate, among other changes.

Finally, we could change the model to consider proportional taxes . In such a case, a new fiscal externality would arise. In fact, any positive effect of UI on wages would compensate the higher expenses. In this sense, by abstracting from proportional taxes we are, again, underestimating the welfare gains whenever $\eta_{w,b}>0$.

We now turn to discuss some aspects of how we implement these formulas in the data.
In our empirical implementation we analyze re-employment wages and the sum of the UI paid for each spell as outcome variables. In the model, the wages that are relevant for the estimation are those that are related to the covered unemployed worker, ie. before the exhaustion of UI. In our empirical implementation we analyze all wages, including those that find a job after exhaustion. We do so to increase the sample size. According to our model the positive effects of UI on wages should be stronger for eligible workers, so while using all our sample we can assume that we are underestimating the LHS of the formula.

When implementing the RKD on wages with controls, we include a set of dummies that capture unemployment duration. This is potentially important, given that changes in the level of benefits would potentially affect also unemployment duration. At the same time, unemployment duration can affect re-employment wages, due to duration dependence. For example, a negative duration dependence would imply that wages tend to fall with unemployment duration; human capital depreciation while unemployed is a possible explanation for this. Under duration dependence, a change in UI level could affect wages indirectly through the change in the proportion of workers re-employed at each duration. Introducing unemployment duration as control mitigates this concern and helps to interpret the effect as a shift in the accepted wage for any duration. In particular, we interpret our results as the change in accepted wages at the beginning of the covered unemployment spell, which is the relevant statistic when the environment is not stationary
\cite{ShimerWerning07}.

In the model, we assume that UI is determined just by a level and a potential duration. The choice of the outcome variable as the duration of covered spell, $B$, or as the total UI\ paid, $R$, would be inconsequential. Nevertheless, in our data, the UI is not constant within the covered spell, and transfers decrease in months 5 and 9. Given that $R$ better captures the effect on the government budget under these changes within the spell, we choose this variable.

\end{document}